\documentclass{aa}
\usepackage{graphicx}
\usepackage{txfonts}

\def\kms{km~s$^{-1}$}
\def\teff{$\rm T_{\rm eff}$}

\def\gr{$\log {\rm g}$}
\def\vt{${\rm v_{\rm t}}$}

\def\ks{${\rm K_{s}}$}

\newlength{\bibitemsep}
\newlength{\bibparskip}
\let\oldthebibliography\thebibliography
\renewcommand\thebibliography[1]{%
  \oldthebibliography{#1}%
  \setlength{\parskip}{\bibitemsep}%
  \setlength{\itemsep}{\bibparskip}%
}

\begin{document}

\title{The chemical DNA of the Magellanic Clouds}
\subtitle{II. High-resolution spectroscopy of the SMC globular clusters \\ NGC~121, NGC~339 and NGC~419
\thanks{Based on observations collected at the ESO-VLT under the programs 
072.D-0507, 073.D-0211, 083.D-0208, 085.D-0375, 086.D-0665, 093.B-0583, 095.D-0290 and 188.B-3002.}}

\author{
A. Mucciarelli\inst{1,2} \and
A. Minelli\inst{1,2} \and
C. Lardo\inst{1} \and
D. Massari\inst{2} \and 
M. Bellazzini\inst{2} \and
D. Romano\inst{2} \and
L. Origlia\inst{2} \and
F. R. Ferraro\inst{1,2}
}

\institute{
Dipartimento  di  Fisica  e  Astronomia  “Augusto  Righi”,  Alma  Mater  Studiorum, Universit\`a  di Bologna, Via Gobetti 93/2, I-40129 Bologna, Italy
\and
INAF - Osservatorio di Astrofisica e Scienza dello Spazio di Bologna, Via Gobetti 93/3, I-40129 Bologna, Italy
}

\authorrunning{A. Mucciarelli et al.}
\titlerunning{SMC globular clusters}

\abstract{

The Small Magellanic Cloud (SMC) is the host of a rich system of globular clusters (GCs) that span a wide age range.  
The chemical composition of the SMC clusters is still poorly understood, despite their significance to chemical evolution studies. 
Here, we provide the first detailed chemical study of evolved giants in three distinct clusters, NGC~121 (10.5 Gyr), NGC~339 (6 Gyr), 
and NGC~419 (1.4 Gyr). Results are based on high-resolution spectra obtained with FLAMES at the Very Large Telescope.

The chemical fingerprints of these clusters closely resemble those of SMC field stars, supporting the SMC's specific history 
of chemical enrichment relative to the Milky Way. The approximately solar-scaled [$\alpha$/Fe] observed in all three clusters, 
independent of their [Fe/H], demonstrate the SMC's low star formation efficiency. Compared to their Milky Way counterparts, 
elements primarily produced by massive stars are severely underrepresented. Particularly, the young cluster NGC~419's 
extremely low [Zn/Fe] shows that hypernovae have contributed relatively little during the past two Gyr. The three GCs have 
high [Eu/Fe] values regardless of their age. This suggests that the production of the r-process elements in the SMC was 
extremely efficient up to 1.5 Gyr ago, with an enrichment timescale comparable to that from Type Ia supernovae . 

When the properties of the oldest SMC object NGC~121 are compared to those of in-situ Milky Way clusters and accreted clusters 
linked to the Gaia-Enceladus merger event, it is shown that the SMC had already attained the same metallicity 
as Gaia-Enceladus but with lower [Fe/H] ratios at the age of NGC~121. 
This suggests that the chemical enrichment histories of the early SMC and Gaia-Enceladus differed, and that the 
SMC probably had a lower mass in its early ages than Gaia-Enceladus.}
 

\maketitle

\section{Introduction} 

The Milky Way's (MW) most massive satellites, the Large and Small Magellanic Clouds (LMC and SMC, respectively), 
offer the chance to study a pair of interacting galaxies before they merge and are cannibalised by the Galaxy.
The currently prevailing view for their formation and evolution is that the two galaxies evolved independently
and only recently formed a binary pair \citep{diaz11,donghia16}.
This binary system is probably at its first peri-Galactic passage \citep{Shuter1992, Byrd1994, Besla2007,Besla2012}. 
The tidal interactions occurring between these galaxies are responsible for the disturbed structure of the SMC, 
as well as of the LMC disc warp and of the complex star formation (SF) history 
of both galaxies, characterised by different, synchronous SF bursts \citep{Harris2009, Rubele2012, Nidever2020,Massana2022}.

In the last decade, there has been a renewed interest in the turbulent evolution of these irregular galaxies and 
their star populations, as evidenced by several dedicated photometric surveys, as VMC \citep[VISTA survey 
of the Magellanic Clouds system,][]{cioni11}, 
STEP \citep[the SMC in Time: Evolution of a Prototype interacting late-type dwarf galaxy,][]{ripepi14},
SMASH \citep[Survey of the MAgellanic Stellar History,][]{nidever17}, 
VISCACHA \citep[VIsible Soar photometry of star Clusters in tApii and Coxi HuguA,][]{maia19}. 
Additionally, they will be privileged targets for forthcoming multi-object spectrographs, 
like 4MOST \citep[see][]{Cioni2019} and MOONS \citep[see][]{Gonzalez2020}.

The study of the chemical composition of SMC stars has received limited attention, 
despite its importance to understand the chemical enrichment history of this galaxy.
Chemical analyses of high-resolution spectra of SMC red giant branch (RGB) stars 
have been presented only recently.
The APOGEE-2 survey measured abundances for $\sim$1000 SMC RGB stars, 
in particular Fe and $\alpha$-elements \citep{Nidever2020}, Al, Ni, Ce \citep{hasselquist21} 
and Mn \citep{fernandes23}.
This dataset shows a quite flat behaviour of [$\alpha$/Fe] ratios in the range
of [Fe/H] between –1.2 and –0.2 dex, and a knee (the metallicity corresponding to the decrease in [$\alpha$/Fe] 
abundance ratios) likely located at [Fe/H]$<$–2.2 dex. All the measured abundance ratios reveal a clear difference with respect 
to the MW stars, pointing to a slower SF efficiency in the SMC with respect to the MW.
\citet{reggiani21} derived the chemical composition of four SMC metal-poor stars ([Fe/H]$<$-2 dex) 
and for two of them measured also the abundance of the r-process element Eu.
They found that the SMC is more enriched in Eu than the MW and that 
elements like Ba and La are produced from this nucleosynthesis channel at these metallicities.
These two stars are the only ones in the SMC where Eu has been measured.
\citet[][hereafter Paper I]{mucciarelli23} measured Na, $\alpha$, iron-peak and s-process elements for 206 SMC stars, 
belonging to three distinct fields around as many globular clusters (GCs).
They found that different regions in the SMC are characterised by different metallicity and radial velocity distributions. 
Some systematic differences in some abundance ratios are identified in different fields. 
This suggests that the chemical enrichment history in the SMC has not
been uniform, with the presence of possible chemically and kinematic distinct substructures.
Also, all the abundance ratios of species produced by massive stars are significantly lower than 
those measured in the MW, indicating a contribution of these stars to the chemical enrichment of the SMC 
lower than in the MW, according to the low star formation rate expected for this galaxy.

Globular clusters are an excellent tool in the reconstruction of the origin and evolution of the stellar
populations in a galaxy as they enable the simultaneous derivation of ages and metallicities and therefore 
the determination of a reliable age-metallicity relation (AMR). 
In particular, at large distances (as for external resolved galaxies), the ages of GCs are more accurate and precise  
then those of field stars that are largely affected by uncertainties arising from reddening and distance.
Therefore, the abundances derived in GCs  provide time-resolved chemical information 
and are complementary to those of the field stars.
The SMC is the only galaxy in the Local Group that has formed and preserved GCs more or less continuously 
over the past 11 Gyr. As opposed to that, the LMC hosts 15 old GCs, coeval to the MW ones \citep{brocato96,olsen98,wagner17}, 
and a populous family of clusters younger than $\sim$3 Gyr, with a lack of GCs over a broad age range -- the so-called Age Gap; 
the few clusters (ESO 121-SC03, KMHK 1592 and KMHK 1762) falling in this age interval have been 
likely accreted by the SMC~\citep{Mackey2006,piatti22,gatto22}. 
Several SMC GCs fill the age range corresponding to the LMC Age Gap.
The metallicities of SMC GCs have been widely investigated, using photometry \citep{Glatt2008a,Glatt2008b,narloch21} 
and low-resolution spectroscopy \citep{DaCosta1998,Parisi2009,Parisi2015,Dias2021,Parisi2022,Dias2022}, 
to explore the global chemical evolution of the galaxy and its AMR. 
The missing pieces of evidence in our current comprehension of the SMC GCs (and of the SMC 
stellar populations in general) are their abundance patterns. 
The only chemical analyses of SMC GCs based on high-resolution spectroscopy available so far are for 
the very young ($\sim$30 Myr) cluster NGC~330 \citep{hill99} and for the old cluster NGC~121 
\citep[][limited to Fe, Na, O, Mg and Al]{Dalessandro2016}. 
Consequently, the investigation of their chemical composition is still entirely unexplored.

This paper is part of a series aimed at deriving the finer chemical details of field and stellar clusters 
in the Magellanic Clouds, providing new clues to the evolution of these galaxies. 
In Paper I, we analysed the properties of 206 stars located in three fields around three GCs.
In this work, we examine the chemical composition of these three SMC GCs (namely, NGC~121, NGC~339, and NGC~419) 
covering a wide age range. The aim is of studying the SMC's chemical enrichment history at various ages and  
to compare the chemical patterns of these clusters with those of field stars derived in Paper I.
In this work we discuss 18 species in order to study different sites/channels of nucleosynthesis. 
We do not discuss the light elements (O, Na, Mg and Al) involved in the CNO-cycle and observed to 
vary from star to star in GCs; this phenomenon is usually 
referred to as multiple populations \citep[see][and references therein]{gratton12,bastian18} and will be 
discussed in a companion paper.
The paper is structured as follows: in Section~\ref{data} we summarise the observations and describe the spectroscopic dataset. 
Sections~\ref{parameters} and~\ref{analysis} present the atmospheric parameters determination and the chemical 
analysis, respectively. We discuss our results in Sections~\ref{iron}, ~\ref{amr},~\ref{abu},~\ref{comparison}.
Finally we summarise our findings and draw our conclusions in Section~\ref{summary}.

\section{Spectroscopic dataset}\label{data}

NGC~121 is the oldest SMC GC, with an age of 10.5 $\pm$ 0.5 Gyr \citep{Glatt2008a} and a metallicity 
of -- 1.28$\pm$0.03 dex \citep{Dalessandro2016}. 
According to the classification proposed by \citet{Parisi2022}, this cluster is associated with the SMC West Halo, 
a part of the SMC moving outward with respect to the main body of the galaxy \citep{dias16} and probably part of the 
more extended Counter Bridge \citep{diaz12}.\\
NGC~339 is a cluster with an age of 6 $\pm$ 0.5 Gyr \citep{Glatt2008b} and 
[Fe/H] = -1.12$\pm$0.10 dex \citep{DaCosta1998} that has been found to be dynamically unevolved based on its blue stragglers spatial distribution \citep{dresbach}.
It is an example of the SMC intermediate-age clusters lacking in the LMC.
This cluster is associated with the Southern Bridge, a second branch of the stellar Magellanic 
Bridge \citep{belokurov17}.
No high-resolution spectroscopic analysis of this cluster is available in the literature.\\
NGC~419 is the youngest among the three observed GCs, with an age of 1.4$\pm$0.2 Gyr and [Fe/H]=-0.67$\pm$0.12 dex 
\citep{Glatt2008b}. \citet{Parisi2022} associated NGC~419 with the main body of the SMC, 
and stars from the Magellanic bridge have been found in its field of view \citep{massari21}.
Also for this cluster, no high-resolution spectroscopic data exist.

The spectra analysed in this work were collected with the multi-object optical spectrograph FLAMES \citep{Pasquini2002} 
mounted at the Very Large Telescope of the European Southern Observatory under the program 086.D-0665 (PI: Mucciarelli). 
The observations with the UVES-FLAMES spectrograph have been performed adopting the Red Arm 580 UVES setup 
(spectral resolution of R=~47000 and spectral coverage $\sim$ 4800 -- 6800 \AA). 
The observations made with the GIRAFFE spectrograph were obtained with the MEDUSA setups 
HR11 (5597 -- 5840 \AA\ and R = 24200) and HR13 (6120 -- 6406 \AA\ and R = 22500). 
The exposure times and the number of individual exposures for each setup and cluster are reported in Table~\ref{tablog}.
In particular, NGC~121 has been observed with 12 exposures of 2700 sec and one of 2200 sec, 
NGC~339 with 14 of 2700 sec and NGC~419 with 10 of 2700 sec.
For each cluster the same targets have been observed during all the exposures.

After the reduction, performed with the dedicated ESO pipelines\footnote{http://www.eso.org/sci/software/pipelines/} 
(including bias subtraction, flat-fielding, wavelength calibration, spectral extraction, and order  merging), 
the individual spectra of each exposure have been cleaned from the sky contribution by subtracting the spectra of some 
close sky regions observed at the same time as the science targets. Subsequently, single exposures of the same target 
have been combined in an individual spectrum for each star. 

\begin{figure}
\centering
\includegraphics[width=\hsize,clip=true]{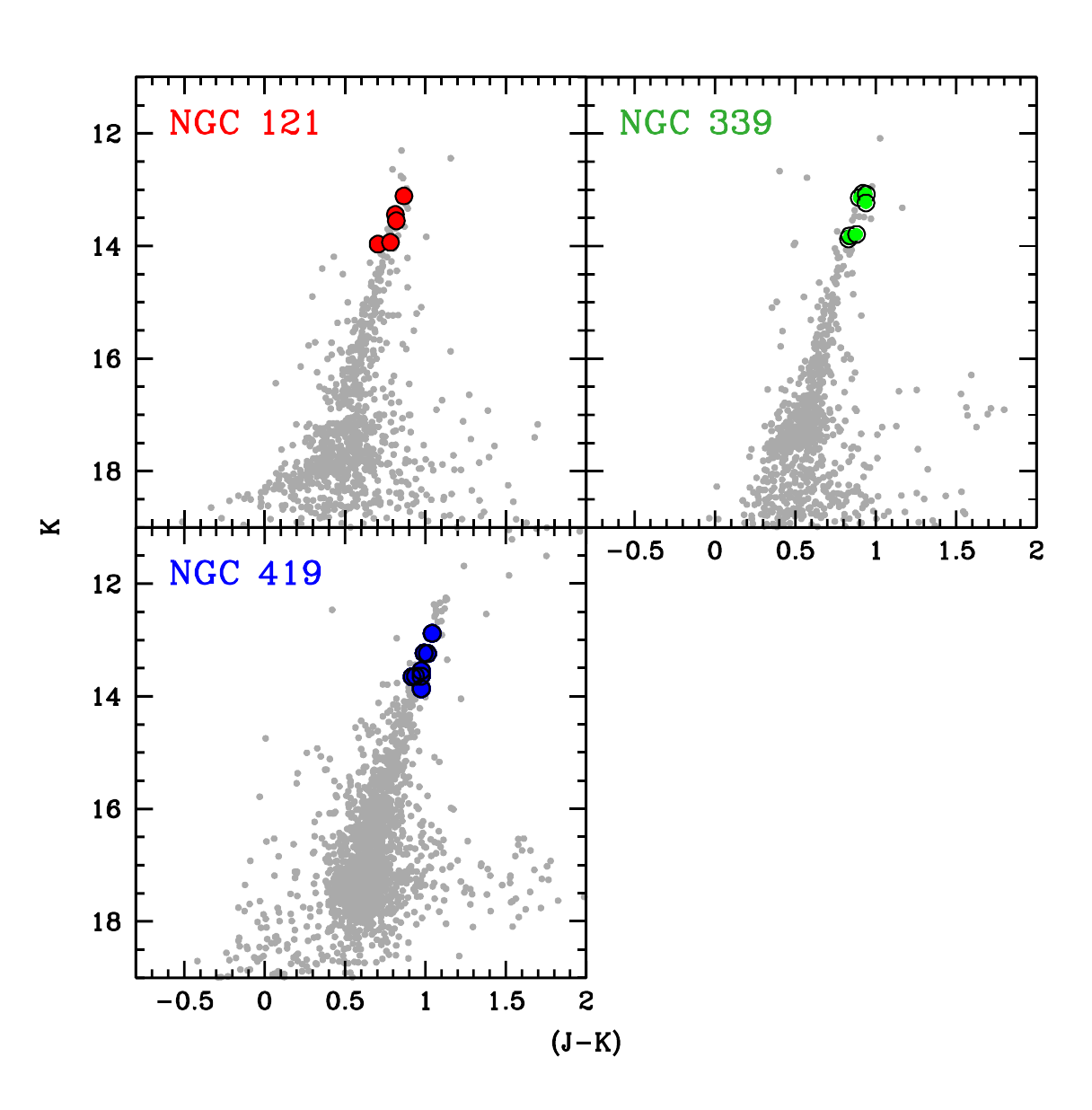}
\caption{(K, J-\ks) colour-magnitude diagrams of the three SMC GCs are plotted as small grey dots. 
Large coloured circles in each panel represent the spectroscopic targets analysed in this work.}
\label{cmd_gc_smc}
\end{figure}

Target stars were selected from the near-infrared photometric catalogues obtained with SofI at the New Technology 
Telescope \citep{m09b} in the brightest portion of the RGB (\ks $\sim$ 13–14). Cluster membership was determined 
using the derived radial velocities (RVs) and metallicities information,  considering the 
targets located within the cluster tidal radius \citep[according to][]{Glatt2009}.
In particular, stars located within the cluster tidal radius with discrepant values 
of RVs and/or [Fe/H] have been considered as SMC field stars and not cluster member stars and 
they have been discussed in Paper I (see Table 2 of Paper I).
The positions in the (K, J-\ks) colour-magnitude diagrams of the cluster member stars 
analysed in this work are shown in Fig.~\ref{cmd_gc_smc}. 

\begin{table*}
\caption{Coordinates of the FLAMES pointing,  number of exposures, and exposure times 
for the GIRAFFE and UVES setups.}
\label{tablog}     
\centering                        
\begin{tabular}{lccccc}        
\hline\hline                 
Field &   RA & Dec  & HR11& HR13  & UVES   \\ \\    
\hline
  &   (J2000)  &  (J2000)  &  &  &    \\
\hline   
NGC~121 & 00:26:49.0   &  --71:32:09.9  &  7x2700sec     &  5x2700sec  &  12x2700 sec \\    
        &              &                &  1x2200sec     &             &   1x2200 sec   \\         
NGC~339 & 00:57:48.9   &  --74:28:00.1  &  9x2700sec     &  5x2700sec  &  14x2700 sec \\  
NGC~419 & 01:08:17.7   &  --72:53:02.7  &  6x2700sec     &  4x2700sec  &  10x2700 sec \\   
\hline
\end{tabular}
\end{table*}

The final sample of cluster stars includes five members of NGC~121 (all observed with UVES), 
seven stars of NGC~339 (four observed with UVES and three with GIRAFFE) and eight stars of NGC~419 
(five observed with UVES and three with GIRAFFE).

Similar to what we did in our previous works \citep[][Paper I]{Minelli2021,Mucciarelli2021}, 
we defined a control sample of MW GCs analysed with the same approach as the SMC targets. 
The MW control sample includes spectra of giant stars of 16 GCs retrieved from the ESO archive and 
obtained with the same UVES-FLAMES configuration (Red Arm 580) used for the majority of target stars of this work. 
The clear advantage of defining a MW control sample is to erase the systematics of the 
analysis when comparing the measured abundances in SMC clusters to those of their MW counterparts. 
These systematics arise mainly from the zero-point of the adopted \teff\ scale, the adopted 
solar reference values, the used atomic data (in particular the log~gf values), the model atmospheres, 
and the code used for the analysis, so that a simple re-scaling to the same solar reference values 
does not suffice to put different datasets on the same scale.

\section{Atmospheric parameters}\label{parameters}

For consistency with Paper I, where the effective temperature (\teff) have been derived 
adopting the colour-\teff\ transformations by \citet{mbm21}, here we used 
$({\rm J-K_{s}})_0$-\teff\ relation by \cite{GonzalezHernandezBonifacio2009}
because these two sets of transformations are based on the same calibration stars.
We used the near-infrared photometry by \citet{m09b}.
Since the colour-\teff\  relation has a mild dependence on stellar metallicity,
we adopted the [Fe/H] values available in the literature as a first step and then we recomputed 
\teff\ with metallicity values determined from the spectroscopic analysis
(anyway leading to variation of less than 20 K with respect to the first run).
Colour excess E(B-V) values are from the reddening maps by \cite{reddening}, 
e.g.,  E(B-V)=0.028, 0.042, and 0.089 mags  for NGC~121, NGC~339, and NGC~419 respectively.
We compared these values with those of the reddening maps by \citet{skowron21} 
finding an excellent agreement: the colour excess values towards NGC~121 and NGC~339 differ 
by less than 0.01 mag, and a difference of 0.03 mag is found for NGC~419. The latter difference 
in E(B-V) leads to variations in \teff\ of about 30 K and in \gr\ of 0.01 dex, with a negligible impact on the 
derived abundances.

Surface gravity (log~g) values are then derived by projecting in \teff\ on
the appropriate isochrone (in terms of age and metallicity) for each GC, computed with the Dartmouth Stellar Evolution Database \citep{Dotter2008}. 
In order to choose the most appropriate isochrone, we first used the age and metallicity determined by \cite{Glatt2008b}, 
refining the metallicity value using those determined in the subsequent analysis. 
The final values chosen for the isochrones are: 10.5 Gyr and [Fe/H] = -- 1.2 dex for NGC~121, 
6 Gyr and [Fe/H] = -- 1.2 dex for NGC~339 and 1.4 Gyr and [Fe/H] = -- 0.6 dex for NGC~419. 
We used solar-scaled isochrones according to our results for the $\alpha$-elements abundances 
(see Section~\ref{abu}).

Finally, the microturbolent velocities \vt\ for UVES spectra have been determined spectroscopically with 
the code {\tt GALA} \citep{Mucciarelli2013},  
minimising the slope between the abundances from Fe I lines and the logarithm of the EWs normalized to 
the wavelength. 
For the GIRAFFE spectra, the spectroscopically \vt\ risk to be affected by uncertainties 
due to the low number of available Fe lines. 
Therefore we compute them from the log~g-\vt\ relation provided by \cite{microturb}. 
We checked for the UVES targets that the two approaches provide consistent results, with an average 
difference between \vt\ from the relation and those obtained from the optimisation process 
of +0.14 \kms\ ($\sigma$=0.13 \kms).
Final atmospheric parameters for the target stars are listed in Table \ref{param_smc_gc}.

\begin{table*}
    \centering
\caption{Adopted atmospheric parameters for SMC GCs targets. 
The ID numbers are referred to SofI photometric catalogs. 
The adopted spectrograph is reported in the last column.}
\label{param_smc_gc}
\begin{tabular}{c c c c c c c c}
\hline
ID   &   RA   &   Dec   &   \teff   &   log g   &   \vt   &   RV   &   spectrum\\
   &   (Degrees)   &   (Degrees)   &   (K)   &      &   (km/s)   &   (km/s)   &   \\
\hline
\multicolumn{8}{c}{NGC~121}\\
\hline
       9   &   6.6842639   &   -71.5367593   &   3990   &   0.53   &   1.60   &   +143.71$\pm$0.05   &   UVES\\
      14   &   6.7033858   &   -71.5295107   &   4070   &   0.66   &   1.80   &   +142.12$\pm$0.05   &   UVES\\
      18   &   6.7233897   &   -71.5450034   &   4110   &   0.73   &   1.70   &   +145.35$\pm$0.05   &   UVES\\
      31   &   6.6845726   &   -71.5315588   &   4250   &   0.97   &   1.60   &   +144.67$\pm$0.07   &   UVES\\
      35   &   6.6970187   &   -71.5477942   &   4240   &   0.95   &   1.40   &   +147.80$\pm$0.07   &   UVES\\
\hline
\multicolumn{8}{c}{NGC~339}\\
\hline
     219   &   14.4805308   &   -74.4371549   &   4140   &   0.72   &   1.80   &   +115.30$\pm$0.14   &   GIRAFFE\\
     466   &   14.4346714   &   -74.4407691   &   4000   &   0.50   &   1.90   &   +112.60$\pm$0.12   &   GIRAFFE\\
     535   &   14.4182781   &   -74.4629265   &   4000   &   0.50   &   1.70   &   +110.83$\pm$0.03   &   UVES\\
     835   &   14.5722246   &   -74.4602041   &   4000   &   0.50   &   1.90   &   +115.70$\pm$0.11   &   GIRAFFE\\
     893   &   14.4515721   &   -74.4772827   &   4050   &   0.58   &   1.60   &   +114.63$\pm$0.03   &   UVES\\
     958   &   14.4013939   &   -74.4746562   &   4210   &   0.83   &   1.50   &   +112.90$\pm$0.04   &   UVES\\
    1076   &   14.4236180   &   -74.4731685   &   4290   &   0.97   &   1.50   &   +111.66$\pm$0.05   &   UVES\\
\hline						 
\multicolumn{8}{c}{NGC~419}\\
\hline
     345   &   17.1136244   &   -72.9015548  & 4320  &  1.41   &   1.56   &	 +189.90$\pm$0.12	     &   GIRAFFE\\ 
     616   &   17.0741376   &   -72.8648011  & 4050  &  0.98   &   1.72   &   +189.30$\pm$0.13		 &   GIRAFFE\\ 
     727   &   17.0584817   &   -72.8773208  & 4270  &  1.33   &   1.60   &   +183.92$\pm$0.05	     &   UVES\\
     732   &   17.0569039   &   -72.8916523  & 4110  &  1.07   &   1.50   &   +186.03$\pm$0.04	     &   UVES\\
     852   &   17.0417327   &   -72.8819103  & 4145  &  1.13   &   1.70   &   +187.37$\pm$0.06	     &   UVES\\
     885   &   17.0363101   &   -72.8775638  & 4190  &  1.20   &   1.64   &	 +191.32$\pm$0.13	     &   GIRAFFE\\ 
    1384   &   17.0854887   &   -72.8729310  & 4150  &  1.13   &   1.70   &   +187.69$\pm$0.03	      &   UVES\\
    1633   &   17.0871666   &   -72.8803339  & 4240  &  1.28   &   1.60   &   +190.41$\pm$0.06	      &   UVES\\
\hline
\hline
\end{tabular}
\end{table*}

\section{Chemical analysis}\label{analysis}

\subsection{Determination of the chemical abundances}

In order to select for the chemical analysis only unblended and linear/poorly saturated lines, 
we compared the observed spectra with synthetic ones computed with the appropriate 
atmospheric parameters and metallicity by using the code { \tt SYNTHE} \citep{Kurucz2005}.
Model atmospheres have been calculated  with the code {\tt ATLAS9} \citep{Kurucz1993,Kurucz2005}. 
As a first guess we assumed for the model atmospheres of all the target clusters an $\alpha$-enhanced 
chemical mixture. The results of the first run of chemical analysis indicated that these stars have 
solar-scaled [$\alpha$/Fe] abundance ratios and we repeated the analysis adopting solar-scaled 
model atmospheres. \\ 
Atomic and molecular data (such as excitation potential $\chi$, log gf, damping constants and 
hyperfine/isotopic splitting) used for synthetic spectra are from the last release of the Kurucz/Castelli 
linelists\footnote{http://wwwuser.oats.inaf.it/castelli/linelists.html}, 
with some exceptions for more recent or more accurate data for some transitions of Fe, Si, Ca, Ti, Ni, Ba 
and Eu \citep[see][for additional references]{Mucciarelli2017}. 
Hyperfine and/or isotopic splitting are taken into account for Sc, V, Mn, Co, Cu, Ba, La, Eu.
The  synthetic spectra have been convolved with a Gaussian profile in order to reproduce the observed 
broadening of GIRAFFE and UVES spectra. 
The final linelists include transitions of elements belonging to the main groups, such as $\alpha$, 
iron-peak and neutron-capture, in particular, 18 elements for UVES spectra and 12 elements for GIRAFFE spectra.

For species with unblended lines (Fe, Ca, Ti, Si, Cr, Ni, Zr, Y and Nd), we measured their EWs 
with {\tt DAOSPEC} \citep{StetsonPancino2008} through the wrapper {\tt 4DAO} \citep{4dao}, 
which also provides RVs of sample stars. Chemical abundances are derived from the measured EWs by using the code {\tt GALA} 
\citep{Mucciarelli2013}. 
The line fitting has been visually inspected line by line, with the purpose to identify possible poorly 
fitted lines or  erroneous determination of the local continuum.\\ 
A different approach is adopted for the species with lines characterised by 
hyperfine/isotopic structure (Sc, V, Mn, Co, Cu, Ba, La, Eu) or transitions located in noisy/complex 
spectral regions (Zn). 
Their abundances have been derived with our own code {\tt SALVADOR} that performs a $\chi^2$-minimisation 
between the observed line and a grid of synthetic models computed with the {\tt SYNTHE} code, 
for which we only vary the abundance of the investigated element.

Solar values are taken from \cite{Grevesse1998} for all the elements. All the abundance ratios for individual stars are listed 
in Tab.~\ref{smc_abu_1} and \ref{smc_abu_2}, together with the corresponding uncertainties as described in Section~\ref{err}.
We adopted a maximum likelihood algorithm \citep[see e.g.][]{pryor93,walker06,Mucciarelli2012} 
to estimate for each cluster the average abundance ratios and their intrinsic scatter together with 
associated uncertainty, obtained by taking into account the uncertainties of abundance ratios in individual stars.
Table~\ref{abu_medie} lists the average abundance ratios for all the targets, together with the intrinsic  
($\sigma_{int}$) and observed ($\sigma_{obs}$) spread.

%
%

\begin{sidewaystable}\tiny
    \centering
\caption{Chemical abundances for the SMC GCs stars, first part.}
\label{smc_abu_1}
\begin{tabular}{ c c ccc ccc ccc }
{\scriptsize ID} & 
{\scriptsize [Fe/H]}& 
{\scriptsize [Fe~II/H]}&
{\scriptsize [Si/Fe]} &
{\scriptsize [Ca/Fe]} &
{\scriptsize [Sc~II/Fe]} &
{\scriptsize [Ti/Fe]} &
{\scriptsize [V/Fe]}  &
{\scriptsize [Cr/Fe]}  &
{\scriptsize [Mn/Fe]}  &
{\scriptsize [Co/Fe]} \\
\hline
\multicolumn{11}{c}{\tiny NGC~121}\\
\hline
       9&-1.22 $\pm$ 0.04 &-1.34 $\pm$ 0.12 & +0.13 $\pm$ 0.08 & +0.02 $\pm$ 0.06 &-0.22 $\pm$ 0.06& +0.05 $\pm$ 0.09 &-0.22 $\pm$ 0.07 &-0.08 $\pm$ 0.14  &-0.50 $\pm$ 0.09  &-0.16 $\pm$ 0.05 \\
      14&-1.21 $\pm$ 0.06 &-1.35 $\pm$ 0.11 & +0.16 $\pm$ 0.09 & +0.05 $\pm$ 0.06 &-0.13 $\pm$ 0.06& +0.02 $\pm$ 0.10 &-0.18 $\pm$ 0.07 &-0.07 $\pm$ 0.15  &-0.55 $\pm$ 0.07  &-0.10 $\pm$ 0.07 \\
      18&-1.13 $\pm$ 0.07 &-1.19 $\pm$ 0.08 & +0.08 $\pm$ 0.10 & +0.06 $\pm$ 0.05 &-0.20 $\pm$ 0.09& +0.03 $\pm$ 0.07 &-0.30 $\pm$ 0.12 &-0.24 $\pm$ 0.10  &-0.63 $\pm$ 0.12  &-0.18 $\pm$ 0.08 \\
      31&-1.10 $\pm$ 0.06 &-1.09 $\pm$ 0.10 & +0.05 $\pm$ 0.12 & +0.08 $\pm$ 0.05 &-0.21 $\pm$ 0.07& +0.07 $\pm$ 0.08 &-0.12 $\pm$ 0.06 &-0.17 $\pm$ 0.13  &-0.52 $\pm$ 0.15  &-0.12 $\pm$ 0.06 \\
      35&-1.18 $\pm$ 0.06 &-1.28 $\pm$ 0.07 & -0.05 $\pm$ 0.11 & +0.01 $\pm$ 0.04 &-0.23 $\pm$ 0.06& +0.04 $\pm$ 0.06 &-0.26 $\pm$ 0.09 &-0.19 $\pm$ 0.14  &-0.64 $\pm$ 0.09  &-0.11 $\pm$ 0.08 \\
\hline
\multicolumn{11}{c}{\tiny NGC~339}\\
\hline
     219&-1.26 $\pm$ 0.11 &--              	& +0.10 $\pm$ 0.12 & +0.18 $\pm$ 0.08 &-- 	            &+0.11 $\pm$ 0.12 &   -0.29 $\pm$ 0.16 &-0.22 $\pm$ 0.12 &--		       & -0.05 $\pm$ 0.09 \\
     466&-1.24 $\pm$ 0.08 &--              	& +0.26 $\pm$ 0.14 & +0.30 $\pm$ 0.10 &-- 	            &+0.10 $\pm$ 0.15 &   -0.17 $\pm$ 0.18 &-0.19 $\pm$ 0.13 &--		       & -0.10 $\pm$ 0.09 \\
     535&-1.24 $\pm$ 0.02 &-1.27 $\pm$ 0.06	& +0.17 $\pm$ 0.04 & +0.05 $\pm$ 0.06 &-0.28 $\pm$ 0.07 &-0.05 $\pm$ 0.04 &   -0.40 $\pm$ 0.06 &-0.18 $\pm$ 0.05 &-0.53 $\pm$ 0.03 & -0.14 $\pm$ 0.03 \\
     835&-1.24 $\pm$ 0.08 &--              	& +0.23 $\pm$ 0.12 & +0.22 $\pm$ 0.12 &-- 	            &+0.05 $\pm$ 0.14 &   -0.35 $\pm$ 0.18 &-0.02 $\pm$ 0.11 &--		       & -0.07 $\pm$ 0.08 \\
     893&-1.25 $\pm$ 0.03 &-1.36 $\pm$ 0.05	& +0.10 $\pm$ 0.05 & +0.11 $\pm$ 0.05 &-0.32 $\pm$ 0.05 &+0.02 $\pm$ 0.05 &   -0.32 $\pm$ 0.07 &-0.11 $\pm$ 0.05 &-0.53 $\pm$ 0.03 & -0.27 $\pm$ 0.03 \\
     958&-1.20 $\pm$ 0.05 &-1.32 $\pm$ 0.05	& +0.18 $\pm$ 0.06 & +0.11 $\pm$ 0.06 &-0.10 $\pm$ 0.09 &+0.04 $\pm$ 0.05 &   -0.22 $\pm$ 0.06 &-0.05 $\pm$ 0.06 &-0.48 $\pm$ 0.09 & -0.04 $\pm$ 0.05 \\
    1076&-1.27 $\pm$ 0.04 &-1.38 $\pm$ 0.05	 & +0.18 $\pm$ 0.05 & +0.16 $\pm$ 0.04 &-0.35 $\pm$ 0.07 &+0.04 $\pm$ 0.03 &   -0.27 $\pm$ 0.07 &-0.13 $\pm$ 0.05 &-0.55 $\pm$ 0.04 & -0.06 $\pm$ 0.04 \\
\hline
\multicolumn{11}{c}{\tiny NGC~419}\\
\hline
     345&-0.54 $\pm$ 0.11   &  --                &-0.08 $\pm$ 0.17  &-0.01 $\pm$ 0.10 &--		        &+0.01 $\pm$ 0.13    &  -0.15 $\pm$ 0.17 &-0.18 $\pm$ 0.15 &-- 	            & -0.19 $\pm$ 0.09  \\
     616&-0.62 $\pm$ 0.09   &  --                &+0.04 $\pm$ 0.13  &+0.01 $\pm$ 0.14 &--		        &+0.01 $\pm$ 0.15    &  -0.32 $\pm$ 0.17 &-0.11 $\pm$ 0.13 &-- 	            & -0.05 $\pm$ 0.09  \\
     727&-0.60 $\pm$ 0.05   &-0.70 $\pm$ 0.07    &+0.02 $\pm$ 0.08  &+0.04 $\pm$ 0.08 &-0.19$\pm$ 0.05   &-0.08 $\pm$ 0.09    &  -0.42$\pm$ 0.11  &-0.10 $\pm$ 0.08 &-0.50$\pm$ 0.08 & -0.30 $\pm$ 0.05 \\
     732&-0.53 $\pm$ 0.05   &-0.67 $\pm$ 0.11    &-0.04 $\pm$ 0.07  &-0.07 $\pm$ 0.10 &-0.32$\pm$ 0.06   &+0.00 $\pm$ 0.10    &  -0.36$\pm$ 0.12  &-0.06 $\pm$ 0.08 &-0.51$\pm$ 0.10 & -0.23 $\pm$ 0.05 \\
     852&-0.61 $\pm$ 0.05   &-0.57 $\pm$ 0.09    &+0.11 $\pm$ 0.08  &-0.09 $\pm$ 0.11 &-0.12$\pm$ 0.07   &-0.22 $\pm$ 0.09    &  -0.52$\pm$ 0.09  &-0.15 $\pm$ 0.08 &-0.54$\pm$ 0.08 & -0.30 $\pm$ 0.05 \\
     885&-0.55 $\pm$ 0.11   &--                  &+0.07 $\pm$ 0.15  &+0.17 $\pm$ 0.14 &--		        &+0.00 $\pm$ 0.14    &  -0.31 $\pm$ 0.17 &-0.08 $\pm$ 0.20 &-- 	            & -0.10 $\pm$ 0.10 \\
    1384&-0.56 $\pm$ 0.04   &-0.67 $\pm$ 0.08    &+0.01 $\pm$ 0.06  &-0.11 $\pm$ 0.08 &-0.22$\pm$ 0.05   &-0.18 $\pm$ 0.08    &  -0.36$\pm$ 0.09  &-0.12 $\pm$ 0.07 &-0.48$\pm$ 0.05 & -0.25 $\pm$ 0.04 \\
    1633&-0.65 $\pm$ 0.05   &-0.72 $\pm$ 0.07    &+0.02 $\pm$ 0.08  &-0.05 $\pm$ 0.07 &-0.26$\pm$ 0.06   &-0.12 $\pm$ 0.08    &  -0.35$\pm$ 0.09  &-0.11 $\pm$ 0.08 &-0.64$\pm$ 0.05 & -0.21 $\pm$ 0.05 \\
\hline
\hline
\end{tabular}
\end{sidewaystable}

\begin{sidewaystable}\tiny
    \centering
\caption{Chemical abundances for the SMC GCs stars, second part.}
\label{smc_abu_2}
\begin{tabular}{ c ccc ccc ccc }
{\scriptsize ID} & 
{\scriptsize [Ni/Fe]}& 
{\scriptsize [Cu/Fe]} &
{\scriptsize [Zn/Fe]} &
{\scriptsize [Y/Fe]} &
{\scriptsize [Zr/Fe]} &
{\scriptsize [Ba~II/Fe]}  &
{\scriptsize [La~II/Fe]}  &
{\scriptsize [Nd~II/Fe]}  &
{\scriptsize [Eu~II/Fe]} \\
\hline
\multicolumn{10}{c}{\tiny NGC~121}\\
\hline
       9  &-0.17 $\pm$ 0.03 &-1.00 $\pm$ 0.22 &-- &-0.42 $\pm$ 0.15 &-0.01 $\pm$ 0.14 &-0.16 $\pm$ 0.0 9&+0.04 $\pm$ 0.08 & +0.50 $\pm$ 0.09 & +0.45 $\pm$ 0.11  \\
      14  &-0.17 $\pm$ 0.04 &-1.13 $\pm$ 0.23 &-- &-0.12 $\pm$ 0.17 &-0.09 $\pm$ 0.16 &-0.10 $\pm$ 0.1 2&+0.03 $\pm$ 0.08 & +0.50 $\pm$ 0.09 & +0.49 $\pm$ 0.12  \\
      18  &-0.11 $\pm$ 0.03 &-0.81 $\pm$ 0.23 &-- &--               &-0.17 $\pm$ 0.14 &-0.01 $\pm$ 0.0 9&-0.05 $\pm$ 0.10 & +0.66 $\pm$ 0.08 & +0.56 $\pm$ 0.12  \\
      31  &-0.20 $\pm$ 0.04 &--               &-- &--               & 0.15 $\pm$ 0.18 &-0.03 $\pm$ 0.1 1&-0.08 $\pm$ 0.08 & +0.57 $\pm$ 0.10 & +0.50 $\pm$ 0.12  \\
      35  &-0.10 $\pm$ 0.04 &--               &-- &--               &-0.05 $\pm$ 0.15 &-0.18 $\pm$ 0.1 0&+0.00 $\pm$ 0.09 & +0.69 $\pm$ 0.09 & +0.37 $\pm$ 0.11  \\
\hline
\multicolumn{10}{c}{\tiny NGC~339}\\
\hline
     2199 &-0.17 $\pm$ 0.05 &-0.67 $\pm$ 0.12 &--               &--               & -0.01 $\pm$ 0.20 & +0.33 $\pm$ 0.09 & +0.07 $\pm$ 0.11 &--	          &--\\
     4669 &-0.22 $\pm$ 0.04 &-0.59 $\pm$ 0.12 &--               &--               & +0.01 $\pm$ 0.21 & +0.16 $\pm$ 0.10 & +0.36 $\pm$ 0.11 &--	          &--\\
     5353 &-0.21 $\pm$ 0.02 &-0.65 $\pm$ 0.07 &-0.24 $\pm$ 0.09 &-0.24 $\pm$ 0.05 & +0.00 $\pm$ 0.08 & +0.24 $\pm$ 0.06 & +0.19 $\pm$ 0.05 & +0.28 $\pm$ 0.04& +0.67 $\pm$ 0.07  \\
     8358 &-0.14 $\pm$ 0.05 &-0.63 $\pm$ 0.11 &--               &--               & -0.09 $\pm$ 0.22 & +0.07 $\pm$ 0.11 & +0.32 $\pm$ 0.11 &--              &--\\
     8933 &-0.22 $\pm$ 0.02 &-0.64 $\pm$ 0.08 &-0.33 $\pm$ 0.09 &-0.22 $\pm$ 0.05 & +0.00 $\pm$ 0.09 & +0.09 $\pm$ 0.06 & +0.17 $\pm$ 0.05 & +0.31 $\pm$ 0.04& +0.63 $\pm$ 0.07  \\
     9585 &-0.17 $\pm$ 0.02 &-0.57 $\pm$ 0.11 &-0.21 $\pm$ 0.14 &-0.09 $\pm$ 0.11 & +0.02 $\pm$ 0.11 & +0.18 $\pm$ 0.09 & +0.20 $\pm$ 0.07 & +0.45 $\pm$ 0.05& +0.71 $\pm$ 0.11  \\
    10764 &-0.15 $\pm$ 0.02 &-0.80 $\pm$ 0.08 &-0.35 $\pm$ 0.10 &-0.27 $\pm$ 0.06 & +0.06 $\pm$ 0.09 & +0.15 $\pm$ 0.07 & +0.18 $\pm$ 0.07 & +0.33 $\pm$ 0.05& +0.72 $\pm$ 0.10  \\
\hline
\multicolumn{10}{c}{\tiny NGC~419}\\
\hline
     345 &-0.07 $\pm$ 0.08 &-0.67 $\pm$ 0.15 &--               &--               & -0.01 $\pm$ 0.20 & +0.27 $\pm$ 0.08 & +0.39 $\pm$ 0.10 &--                &--\\
     616 &-0.15 $\pm$ 0.07 &-0.65 $\pm$ 0.10 &--               &--               & +0.03 $\pm$ 0.20 & +0.28 $\pm$ 0.07 & +0.25 $\pm$ 0.09 &--                &--\\
     727 &-0.22 $\pm$ 0.03 &-0.88 $\pm$ 0.15 &-0.66 $\pm$ 0.15 &-0.11 $\pm$ 0.08 & +0.09 $\pm$ 0.16 & +0.28 $\pm$ 0.08 & +0.29 $\pm$ 0.07 & +0.42 $\pm$ 0.06 & +0.54 $\pm$ 0.10  \\
     732 &-0.28 $\pm$ 0.03 &--  	            &-0.70 $\pm$ 0.15 &-0.33 $\pm$ 0.07 & +0.20 $\pm$ 0.16 & +0.34 $\pm$ 0.06 & +0.35 $\pm$ 0.07 & +0.31 $\pm$ 0.06 & +0.46 $\pm$ 0.07  \\
     852 &-0.15 $\pm$ 0.04 &-0.99 $\pm$ 0.32 &-0.69 $\pm$ 0.19 &-0.16 $\pm$ 0.08 & -0.10 $\pm$ 0.15 & +0.29 $\pm$ 0.13 & +0.26 $\pm$ 0.09 & +0.35 $\pm$ 0.06 & +0.53 $\pm$ 0.12  \\
     885 &-0.25 $\pm$ 0.08 &-0.87 $\pm$ 0.15 &--               &--               & -0.02 $\pm$ 0.20 & +0.20 $\pm$ 0.09 & +0.23 $\pm$ 0.11 &--                &--\\
    1384 &-0.26 $\pm$ 0.03 &-0.83 $\pm$ 0.15 &-0.66 $\pm$ 0.15 &-0.30 $\pm$ 0.06 & -0.02 $\pm$ 0.13 & +0.34 $\pm$ 0.06 & +0.24 $\pm$ 0.06 & +0.27 $\pm$ 0.05 & +0.57 $\pm$ 0.07  \\
    1633 &-0.22 $\pm$ 0.03 &--  	         &--	           &-0.21 $\pm$ 0.08 & +0.06 $\pm$ 0.16 & +0.21 $\pm$ 0.10 & +0.30 $\pm$ 0.07 & +0.27 $\pm$ 0.08 & +0.44 $\pm$ 0.10  \\\hline
\hline
\end{tabular}
\end{sidewaystable}

\begin{table*}\tiny
\caption{Average abundance ratios for the SMC GCs, together 
with the error of the mean, the typical uncertainty of individual stars and the number of stars.}
\label{abu_medie}
\begin{tabular}{c cc cc cc cc cc cc}
&\multicolumn{4}{c}{\footnotesize NGC~121}&\multicolumn{4}{c}{\footnotesize NGC~339}&\multicolumn{4}{c}{\footnotesize NGC~419}\\
&  $< >$  & $\sigma_{int}$ & $\sigma_{obs}$ &N$_{\star}$ 
&  $< >$  & $\sigma_{int}$ & $\sigma_{obs}$ &N$_{\star}$ 
&  $< >$  & $\sigma_{int}$ & $\sigma_{obs}$ &N$_{\star}$ \\
\hline 
$[Fe/H]$  &-1.18 $\pm$ 0.02& 0.00 $\pm$ 0.06   & 0.05 &5  &-1.24 $\pm$ 0.01& 0.00$\pm$  0.02 &  0.02  &7&-0.58 $\pm$ 0.02& 0.00  $\pm$  0.03 & 0.04 & 8\\	      
$[FeII/H]$&-1.24 $\pm$ 0.04& 0.00 $\pm$ 0.09   & 0.11 &5  &-1.34 $\pm$ 0.03& 0.00$\pm$  0.04 &  0.05  &4&-0.68 $\pm$ 0.04& 0.00  $\pm$  0.05 & 0.06 & 5\\	      
$[Si/Fe] $&+0.09 $\pm$ 0.04& 0.00 $\pm$ 0.06   & 0.08 &5  & +0.16 $\pm$ 0.02& 0.00$\pm$  0.03 &  0.06  &7&+0.02 $\pm$ 0.03& 0.00  $\pm$  0.04 & 0.06 & 8\\		      
$[Ca/Fe] $&+0.04 $\pm$ 0.02& 0.00 $\pm$ 0.03   & 0.03 &5  & +0.14 $\pm$ 0.02& 0.00$\pm$  0.04 &  0.08  &7&-0.03 $\pm$ 0.03& 0.00  $\pm$  0.05 & 0.09 & 8\\		      
$[Sc~II/Fe]$&-0.20 $\pm$ 0.03& 0.00 $\pm$ 0.04   & 0.04 &5  & -0.29 $\pm$ 0.03& 0.00$\pm$  0.10 &  0.11  &7&-0.22 $\pm$ 0.03& 0.00  $\pm$  0.10 & 0.07 & 8\\		      
$[Ti/Fe] $&+0.04 $\pm$ 0.03& 0.00 $\pm$ 0.03   & 0.02 &5  &-0.01 $\pm$ 0.02& 0.00$\pm$  0.10 &  0.05  &4&-0.10 $\pm$ 0.04& 0.00  $\pm$  0.06 & 0.09 & 5\\		      
$[V/Fe]  $&-0.20 $\pm$ 0.03& 0.00 $\pm$ 0.06   & 0.07 &5  &-0.30 $\pm$ 0.03& 0.03$\pm$  0.05 &  0.08  &7&-0.38 $\pm$ 0.04& 0.00  $\pm$  0.07 & 0.10 & 8\\		      
$[Cr/Fe] $&-0.17 $\pm$ 0.06& 0.00 $\pm$ 0.07   & 0.07 &5  &-0.13 $\pm$ 0.02& 0.00$\pm$  0.04 &  0.07  &7&-0.11 $\pm$ 0.03& 0.00  $\pm$  0.03 & 0.04 & 8\\		      
$[Mn/Fe] $&-0.57 $\pm$ 0.04& 0.00 $\pm$ 0.05   & 0.06 &5  &-0.53 $\pm$ 0.02& 0.00$\pm$  0.02 &  0.03  &4&-0.54 $\pm$ 0.03& 0.04  $\pm$  0.04 & 0.06 & 5\\		      
$[Co/Fe] $&-0.13 $\pm$ 0.03& 0.00 $\pm$ 0.03   & 0.03 &5  &-0.10 $\pm$ 0.02& 0.00$\pm$  0.08 &  0.04  &7&-0.23 $\pm$ 0.02& 0.00  $\pm$  0.06 & 0.09 & 8\\		      
$[Ni/Fe] $&-0.15 $\pm$ 0.02& 0.01 $\pm$ 0.04   & 0.04 &5  &-0.18 $\pm$ 0.01& 0.02$\pm$  0.01 &  0.03  &7&-0.22 $\pm$ 0.02& 0.03  $\pm$  0.02 & 0.07 & 8\\		      
$[Cu/Fe] $&-0.98 $\pm$ 0.13& 0.00 $\pm$ 0.16   & 0.16 &3  &-0.66 $\pm$ 0.03& 0.00$\pm$  0.06 &  0.07  &7&-0.76 $\pm$ 0.06& 0.00  $\pm$  0.11 & 0.13 & 6\\		      
$[Zn/Fe] $&   ---           & 	 ---	       & ---  &---&-0.29 $\pm$ 0.05& 0.00$\pm$  0.06 &  0.07  &4&-0.68 $\pm$ 0.08& 0.00  $\pm$  0.08 & 0.02 & 4\\		   
$[Y/Fe]  $&-0.29 $\pm$ 0.11& 0.00 $\pm$ 0.30   & 0.21 &2  &+0.23 $\pm$ 0.03& 0.00$\pm$  0.03 &  0.08  &4&-0.23 $\pm$ 0.04& 0.04  $\pm$  0.05 & 0.09 & 5\\		      
$[Zr/Fe] $&-0.04 $\pm$ 0.07& 0.00 $\pm$ 0.09   & 0.12 &5  & +0.01 $\pm$ 0.04& 0.00$\pm$  0.04 &  0.05  &7&+0.03 $\pm$ 0.06& 0.00  $\pm$  0.07 & 0.09 & 8\\		      
$[Ba~II/Fe] $&-0.10 $\pm$ 0.04& 0.00 $\pm$ 0.07   & 0.08 &5  & +0.17 $\pm$ 0.03& 0.02$\pm$  0.08 &  0.09  &7& +0.30 $\pm$ 0.03& 0.00  $\pm$  0.04 & 0.05 & 8\\		      
$[La~II/Fe] $&-0.01 $\pm$ 0.04& 0.00 $\pm$ 0.05   & 0.05 &5  & +0.19 $\pm$ 0.03& 0.00$\pm$  0.03 &  0.10  &7& +0.29 $\pm$ 0.03& 0.00  $\pm$  0.04 & 0.06 & 8\\		      
$[Nd~II/Fe] $   &+0.58 $\pm$ 0.04& 0.00 $\pm$ 0.10   & 0.09 &5  & +0.33 $\pm$ 0.03& 0.04$\pm$  0.04 &  0.07  &4& +0.32 $\pm$ 0.03& 0.00  $\pm$  0.12 & 0.06 & 5\\		       
$[Eu~II/Fe] $&+0.47 $\pm$ 0.05& 0.00 $\pm$ 0.06   & 0.07 &5  & +0.67 $\pm$ 0.04& 0.00$\pm$  0.05 &  0.04  &4& +0.51 $\pm$ 0.04& 0.00  $\pm$  0.05 & 0.06 & 5\\

\hline							  
\hline							  
\end{tabular}
\end{table*}

\normalsize

\subsection{Error estimates}
\label{err}
Abundance uncertainties have been computed combining in quadrature the errors arising from the measure of lines strength 
and the uncertainties arising from atmospheric parameters (see Paper I, for a detailed description). 
Internal errors due to the measurements have been estimated as the standard error of the mean of the 
abundances of individual stars. For abundances from only one line, we propagated the error in the EW 
as provided by DAOSPEC or, for abundances from spectral synthesis, we rely on Montecarlo simulations 
\citep[see][for details]{Minelli2021}.
The uncertainties arising from the atmospheric parameters have been computed deriving the abundance variation due 
to the change of only one parameter at a time, keeping the others fixed, except for~\teff\, which affects the values 
derived for log g and \vt\ that had to be changed accordingly.

The log~g value depends on \teff\ and on the isochrone adopted. Uncertainties related to the choice of a specific isochrone 
are derived from the  age and metallicity values adopted for each cluster. In particular, we compute the variation of log~g 
determined by a change of 1 Gyr in age and 0.1 dex in metallicity, which are the typical errors associated with these parameters. 
Specifically, variations in metallicity mainly affect the typical error associated with log~g measurements,
which is of the order of 0.07 dex.

Finally, typical errors associated with~\vt\ measurements are again different for GIRAFFE and UVES spectra, depending 
on the method used to derive their values. For UVES spectra, \vt\ is derived spectroscopically and the typical uncertainties 
are of the order of 0.1 \kms .
For GIRAFFE targets \vt\ values have been derived from the log~g-\vt\ relation by \citet{microturb} and the 
adopted uncertainty is of 0.15 \kms\ , taking into account the errors in log~g and in the adopted calibration.

\section{[Fe/H] abundances and RVs}\label{iron}
\label{feab}
Here we discuss iron abundances and the systemic RVs for the three target clusters, 
comparing them with estimates from literature . For each cluster we provide the average 
[Fe/H] and RV together with the standard error of the average and the observed standard deviation 
($\sigma_{obs}$).

For NGC~121 we derive an iron abundance of [Fe/H] = --1.18$\pm$0.02 ($\sigma_{obs}$=0.05) dex.  
\citet{DaCosta1998} derived for this cluster [Fe/H]=--1.19$\pm$0.12 dex using Ca~II triplet.
The same high-resolution dataset discussed here has been previously analysed by \citet{Dalessandro2016}. 
They find an average [Fe/H] of -- 1.28$\pm$0.03 dex ($\sigma$=~0.06 dex). 
The small difference between the two values is due to the differences in \teff\ . 
Indeed, the average difference between the \teff\ values derived in this study and those by \citet{Dalessandro2016} 
is +122 K, reflecting the difference in the zero-point of the adopted \teff\ scales --
\citet{alonso99} in \citet{Dalessandro2016} and \citet{GonzalezHernandezBonifacio2009} in this study. 
The mean RV of this cluster is +144.7$\pm$0.9 \kms\ ($\sigma_{obs}$=~1.9 \kms ), in good agreement 
with previous estimates by low-resolution spectra. 
RVs measured from integrated spectra of NGC~121 provide +139$\pm$20 \kms \citep{zinn}, 
+138$\pm$15 \kms\ \citep{hesser} and +147$\pm$2 \kms\ \citep{dubath}, while measures based on 
Ca~II triplet lines of individual stars provide +138$\pm$4 \kms\ \citep{DaCosta1998}.

The intermediate-age GC NGC~339 has a metallicity similar to that of NGC~121, with 
[Fe/H] = -- 1.24$\pm$0.01 ($\sigma_{obs}$=0.02) dex. This value is consistent with the ones obtained 
from the Ca~II triplet, [Fe/H]=--1.19$\pm$0.10 dex \citep{DaCosta1998}  and photometry, 
[Fe/H]=--1.10$\pm$0.12 dex \citep{narloch21}.
The mean RV of NGC~339 is +113.4$\pm$0.7 \kms\ ($\sigma_{obs}$=~1.9 \kms ) in excellent agreement 
with the value by \citet{song2021} of +112.9$\pm$0.4 \kms\ obtained from high-resolution  
spectra collected with M2FS on the Magellan/Clay Telescope. On the other hand, the average RV 
provided by \citet{Parisi2022} from low-resolution spectra is $\sim$10 km/s lower than ours.
They attribute the similar difference with respect to the \citet{song2021} estimate to mis-centering 
of their stars in the slit.

Finally, the intermediate-age GC NGC~419 has a metallicity significantly higher than the other two clusters 
([Fe/H] =--0.58$\pm$0.02 dex, $\sigma_{obs}$=0.04 dex). For this cluster \citet{Parisi2022} 
provided [Fe/H]=--0.62$\pm$0.02 dex from Ca~II triplet.
Its mean RV is +188.2$\pm$0.9 \kms\ ($\sigma_{obs}$=~2.5 \kms ). Similar to NGC~339, we have a good agreement 
with the measure by \citet{song2021} providing 189.9$\pm$0.3 \kms\ , while the value by \citet{Parisi2022} 
is lower by $\sim$17 \kms\ .

\section{Age-metallicity relation}
\label{amr}

\begin{figure}
\includegraphics[width=\hsize,clip=true]{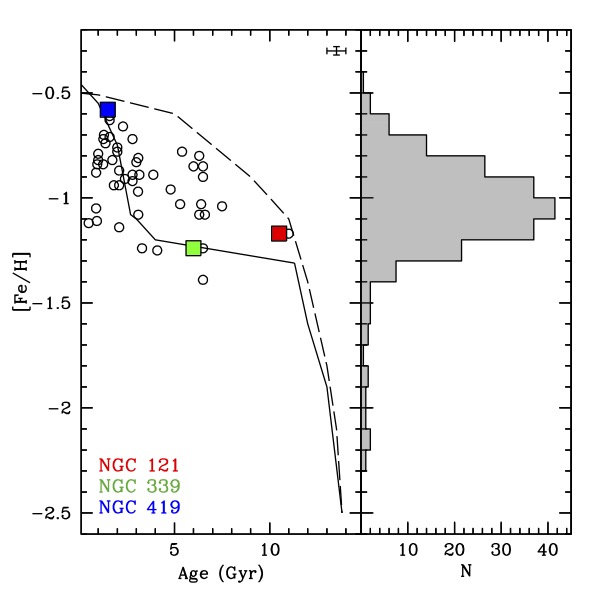}
\caption{Age-metallicity relation of the SMC clusters.
Left panel: average metallicity as a function of the age \citep{Glatt2009} 
for the three SMC GCs analysed in this work (same colour code of Fig.~\ref{cmd_gc_smc}). 
SMC GCs from previous low-resolution works \citep{Parisi2009,Parisi2015,Dias2021,Parisi2022} 
are plotted as open circles. Theoretical AMRs calculated by \cite{Pagel1998} are also reported 
(the solid curve is the bursting model and the dashed line the continuous model).
Right panel: metallicity distribution of the SMC field stars (Paper I).}
\label{age_met_rel}
\end{figure}

The left panel of Fig.~\ref{age_met_rel} shows the behaviour of the measured [Fe/H] for the three target GCs 
as a function of the age. Age estimates are from \citet{Glatt2008a,Glatt2008b}. Thanks to NGC 121, the metallicity reached 
by the SMC $\sim$2--3 Gyr after the earliest SF bursts may be precisely determined. 
Its metallicity ([Fe/H]=--1.18 dex) is lower than that reached by the 
LMC \citep{Pagel1998, Harris2009} and MW \citep{Haywood2013, Snaith2015} at the same epoch, as expected for 
a galaxy characterised by a lower SF efficiency. 
In Section~\ref{comparison} we compare in details the properties of NGC~121 with the empirical 
AMR defined by the MW clusters.

The [Fe/H] distribution of the SMC field stars analysed in Paper I is displayed in the right panel of Fig. ~\ref{age_met_rel}. 
The comparison between this [Fe/H] distribution and the metallicity reached by the SMC at the different ages of the 
three target GCs allows us to constrain the age of the most metal-poor SMC field stars. 
In fact, it is likely that SMC field stars with [Fe/H] $<$ --1.4/--1.5 dex (considerably lower than that of NGC~121) 
were created in the early epochs of the galaxy, showing its early enrichment \citep[][Paper I]{Nidever2020,reggiani21}.

Despite the large difference in their ages ($\sim$4-5 Gyr), NGC~121 and NGC~339 have similar [Fe/H], 
hinting to a very low SF efficiency at those epochs that leads to an almost constant [Fe/H] 
over a large age range, or, perhaps, the effects of outflows. 
However, it is worth noting that the population of SMC GCs coeval to NGC~339 exhibits 
a significant spread in [Fe/H] as estimated using the CaII triplet strength \citep{Parisi2009,Parisi2015,Dias2021,Parisi2022}. 
This [Fe/H] dispersion can be only partially explained by the metallicity gradient inside the SMC 
\citep[see e.g.][Paper I]{Carrera2008,Parisi2016,grady21}, and it may be indicative that 
different regions of the SMC could be characterised by different chemical enrichment histories 
and AMR \citep[as already suggested by][and Paper I]{dias16}.
The metallicity of NGC~339 only is not sufficient to disentangle the origin of the  
metallicity spread at those ages, but it could suggest a low SF efficiency in the 
Southern Bridge (the region where NGC~339 is located) with respect to other 
regions of the SMC.

Finally, NGC~419 has a significant higher [Fe/H] abundance likely due to the sequence of intense bursts 
of SF occurring in the last 3 Gyr \citep{Harris2009,Rubele2012,Massana2022}, 
confirming that the chemical enrichment of the SMC continues at a quicker pace in the last Gyr, 
increasing [Fe/H] up to $\sim$--0.6 dex.

In Fig.~\ref{age_met_rel} we compare our results with the two theoretical AMR trends computed by \cite{Pagel1998}, 
the first one assuming a closed-box model with a continuous SF rate and the second one with a long period 
without SF (or characterised by a very low-level activity) between 12 and 4 Gyr ago and a SF burst 4 Gyr ago. 
Other AMR have been proposed in the literature, 
both theoretical \citep{tb09} and based on observed SF histories \citep{Harris2004,cignoni13} 
and that are consistent with or intermediate between the two models by \citet{Pagel1998}.
The two models by \citet{Pagel1998} are similar with each other at the ages of NGC~121 and NGC~419 
while they clearly disagree at intermediate ages (corresponding to the age of NGC339 and where a 
large [Fe/H] scatter among GCs has been observed). This highlights the importance to derive 
with high-resolution spectroscopy the chemical composition of other clusters coeval to NGC~339.

\section{Abundance ratios}
\label{abu}
We discuss the abundance ratios of the three target GCs for 
$\alpha$ (Si, Ca, Ti), iron-peak (Sc, V, Cr, Mn, Co, Ni, Cu, Zn), 
s-process (Y, Zr, Ba, La, Nd) and r-process (Eu) elements. 
For all the abundances ratios discussed in this section, the observed scatters are compatible 
with null intrinsic scatter (see Table~\ref{abu_medie}). 
Figs. \ref{alfa_el_smc} - \ref{neutron_el_smc} show the individual abundance ratios as a function of [Fe/H] 
for the SMC GCs, the MW GCs of the control sample and the SMC field stars from Paper I.  
It is worth noting that we consider the stars of Paper I regardless of their position within the galaxy. 
However, differences in the metallicity distribution of different regions, as well as some small systematic differences 
in the abundance patterns of some elements have been detected, enforcing the idea that the SMC has been 
characterised by a spatially non-uniform chemical enrichment history.
In addition, also literature abundances from high-resolution spectra for MW field stars are shown as reference.
Mean abundance ratios for the MW control sample are listed in Appendix~\ref{app}.

\subsection{$\alpha$-elements}
Fig. \ref{alfa_el_smc} shows the abundance ratios of the explosive $\alpha$-elements Si, Ca, Ti and 
their average abundance. These elements are mainly produced in massive stars 
and ejected in the interstellar medium via CC-SNe, with a small but 
not negligible contribution by SNe~Ia, especially for Ca and Ti \citep[see e.g.][]{Kobayashi2020}. 

The abundance ratios for the SMC GCs nicely agree with those of the SMC field stars. All the clusters 
display [$\alpha$/Fe] lower than those measured in MW (field and GCs) stars of similar metallicity. 
In particular, NGC~121 has nearly solar-scaled [$\alpha$/Fe] abundance ratios, pointing out that at that age 
($\sim$10.5 Gyr) the SMC has been already enriched by SN~Ia 
(see also Section~\ref{comparison}).

NGC~339 shows only marginally higher (by $\sim$0.1 dex) [Si/Fe] and [Ca/Fe] values and similar [Ti/Fe]. 
We can consider that these two clusters have similar [$\alpha$/Fe]. 
This is evidence for a poor enrichment both in Fe and $\alpha$-elements  over the large range of time 
covered by these two clusters, thus pointing to a low SF rate 
when the SMC evolved in isolation.
Lower [$\alpha$/Fe] values are found in NGC~419, demonstrating that also at that age 
the chemical enrichment is dominated by SN~Ia.
Overall, the run of [$\alpha$/Fe] with [Fe/H] for these three GCs agrees with the results 
from the APOGEE survey \citep{Nidever2020,hasselquist21,fernandes23}.

\begin{figure*}
\centering
\includegraphics[scale=0.6]{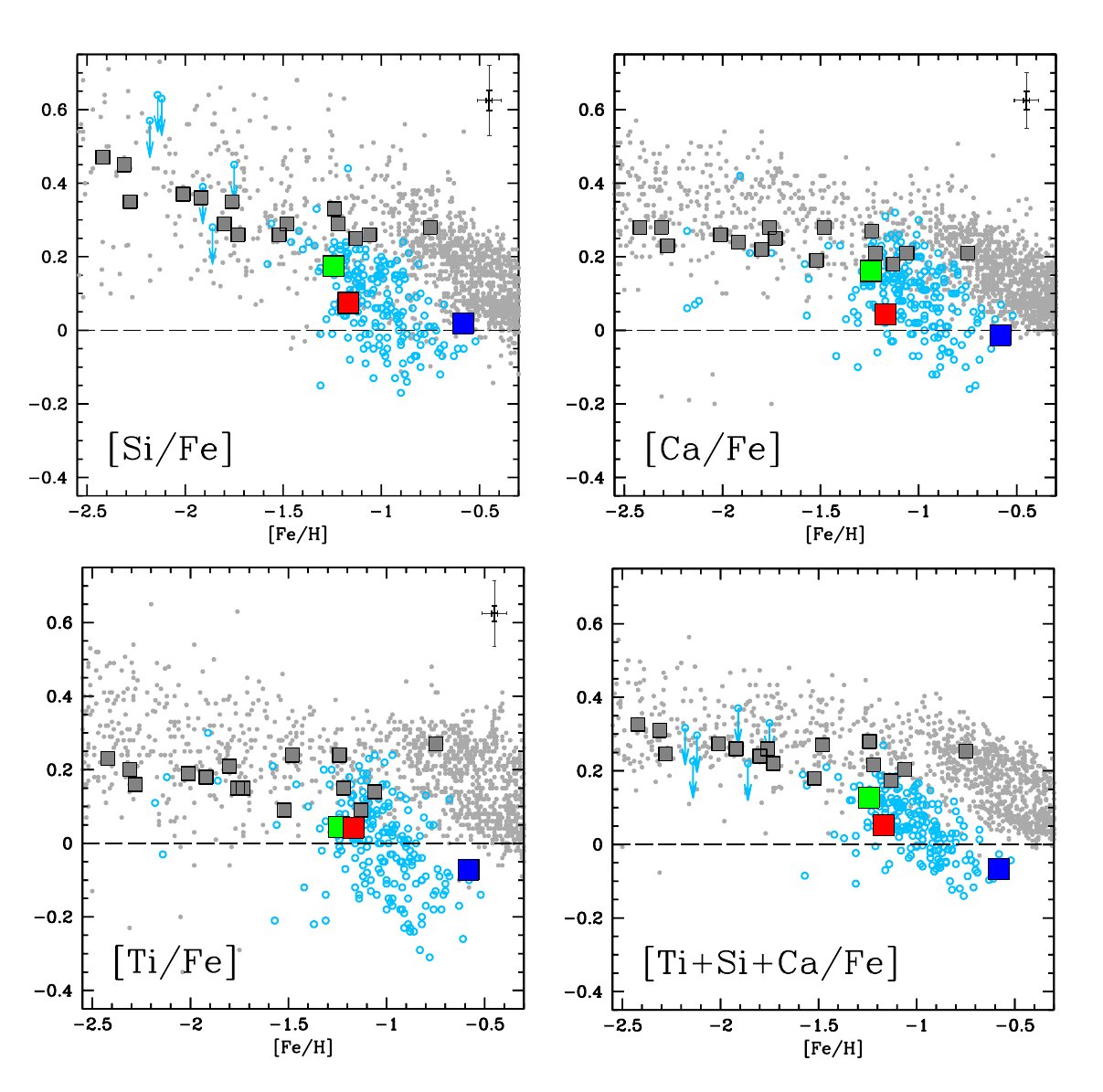}
\caption{$\alpha$-elements abundance ratios (Si, Ca, Ti and their mean value, from top left panel 
to bottom right panel, respectively) as a function of [Fe/H] for SMC GCs (coloured squares: NGC~121 in red, 
NGC~339 in green and NGC~419 in blue), SMC field stars \citep[light blue open dots,][]{Mucciarelli2021}, 
MW GCs from the control sample (grey squares) and MW field stars (grey dots). Arrows indicate upper limits.  
MW field stars are from \cite{Edvardsson1993, Fulbright2000, Stephens2002, Gratton2003, 
Reddy2003, Reddy2006, Barklem2005, Bensby2005, Adibekyan2012, Roederer2012, mishenina2013, Reggiani2017}.
The abundance errors for the SMC GCs are reported as error bars, where the thin bar is the typical total 
uncertainty for individual stars and the thick bar is the error of the mean value of the GC.}
\label{alfa_el_smc}
\end{figure*}

\subsection{Iron-peak elements}

Iron-peak elements are mainly produced in CC-SNe (both standard Type II and HNe) with an important 
contribution by SN~Ia for some of them. In particular, Sc, Cu and Zn are almost completely produced by massive stars, 
V and Co are mainly produced by massive stars but with a small contribution by SN~Ia while most of Cr, Mn 
and Ni are from SN~Ia \citep{romano10,Kobayashi2020,palla2020}. 
We stress that abundances of these elements in the SMC are still poorly investigated: apart from the Sc, V, Ni 
and Cu abundances we derived in Paper I, only abundances of Ni for SMC field stars obtained by the 
APOGEE survey has been discussed \citep{hasselquist21}.

Almost all the iron-peak elements measured in these GCs exhibit abundance ratios lower than those observed in MW stars. 
In particular, these differences are more pronounced in those elements produced mainly in massive stars, like Sc, V, and Zn. 
NGC 419 generally shows the most pronounced discrepancies, 
with differences reaching --0.8 and --0.7 dex for [Cu/Fe] and [Zn/Fe], respectively. 
In the following we discuss in details the abundances of three key iron-peak elements that 
are crucial to constrain the chemical enrichment history of the SMC, namely 
Mn (produced almost totally in SN~Ia), Cu (produced via s-process in massive stars) and 
Zn (almost totally built by HNe).

In our Galaxy [Mn/Fe] is sub-solar until [Fe/H]$\sim$--1 dex because at these metallicities
it is produced only by CC-SNe and in a small amount. At higher metallicities, [Mn/Fe] significantly increases 
due to the dominant contribution by SNe~Ia. The Mn yields increase with 
the metallicity of the SN Ia progenitor and are also sensitive to the explosion mechanism 
of the SNe-Ia \citep{badenes08}. This behaviour is well reproduced by our MW control sample 
\citep[see also][]{sobeck06}. Concerning the SMC, NGC~121 and NGC~339 have [Mn/Fe] values 
compatible with those of the MW control sample, indicating a similar low production of Mn 
at early times. 
At higher metallicities, NGC~419 has a [Mn/Fe] comparable with the other two GCs and in stark 
disagreement with the MW stars of similar [Fe/H].
This suggests that at these ages the enrichment of Mn has been dominated by metal-poor SN~Ia 
(or at least more-metal poor than those dominating the Mn production in our Galaxy).

Copper is mainly produced in massive stars via the weak s-process.
In our Galaxy [Cu/Fe] increases with the metallicity reflecting the metallicity 
dependence of the Cu yields \citep{romano07}. As discussed in Paper I, the SMC field stars exhibit [Cu/Fe] 
lower than the MW stars, despite the large star-to-star scatter limits our conclusions 
about the average trend of [Cu/Fe] with [Fe/H]. Concerning the SMC clusters, 
NGC~121 and NGC~419 have [Cu/Fe] significantly lower than the MW stars and GCs, while 
NGC~339 has a value barely consistent with the MW control sample but still 
compatible also with the SMC field stars.
In general, the low [Cu/Fe] values in these clusters, in particular in NGC~419, 
support a scenario where the contribution by massive stars is significantly reduced 
in the SMC.

Zinc is produced mainly in HNe, explosions with energy significantly larger than 
those of standard CC-SNe and related to stars more massive than $\sim$30-35 $M_{\odot}$ 
\citep{kob06,nomoto13}. 
Therefore, Zn offers crucial hints regarding the contribution of HNe and high-mass stars.
We are able to measure Zn only in NGC~339 and NGC~419, both having [Zn/Fe] values 
lower than the MW stars. In particular, NGC~419 has [Zn/Fe]$\simeq$--0.7 dex, while 
the MW stars at similar metallicities have nearly solar abundance ratios. 
Also this element supports a chemical enrichment in the SMC where the contribution by 
massive stars (and in particular by HNe) is significantly reduced with respect to the MW. 
Low [Zn/Fe] abundance ratios have been measured in other dwarf galaxies characterised by a 
low SF efficiency and a small contribution by massive stars, like Sagittarius 
\citep{sbordone07, Minelli2021}, Sculptor \citep{skul} and the LMC \citep{Mucciarelli2021}, 
confirming that systems less massive than the MW have a lower contribution by HNe.

\begin{figure*}
\centering
\includegraphics[scale=0.45]{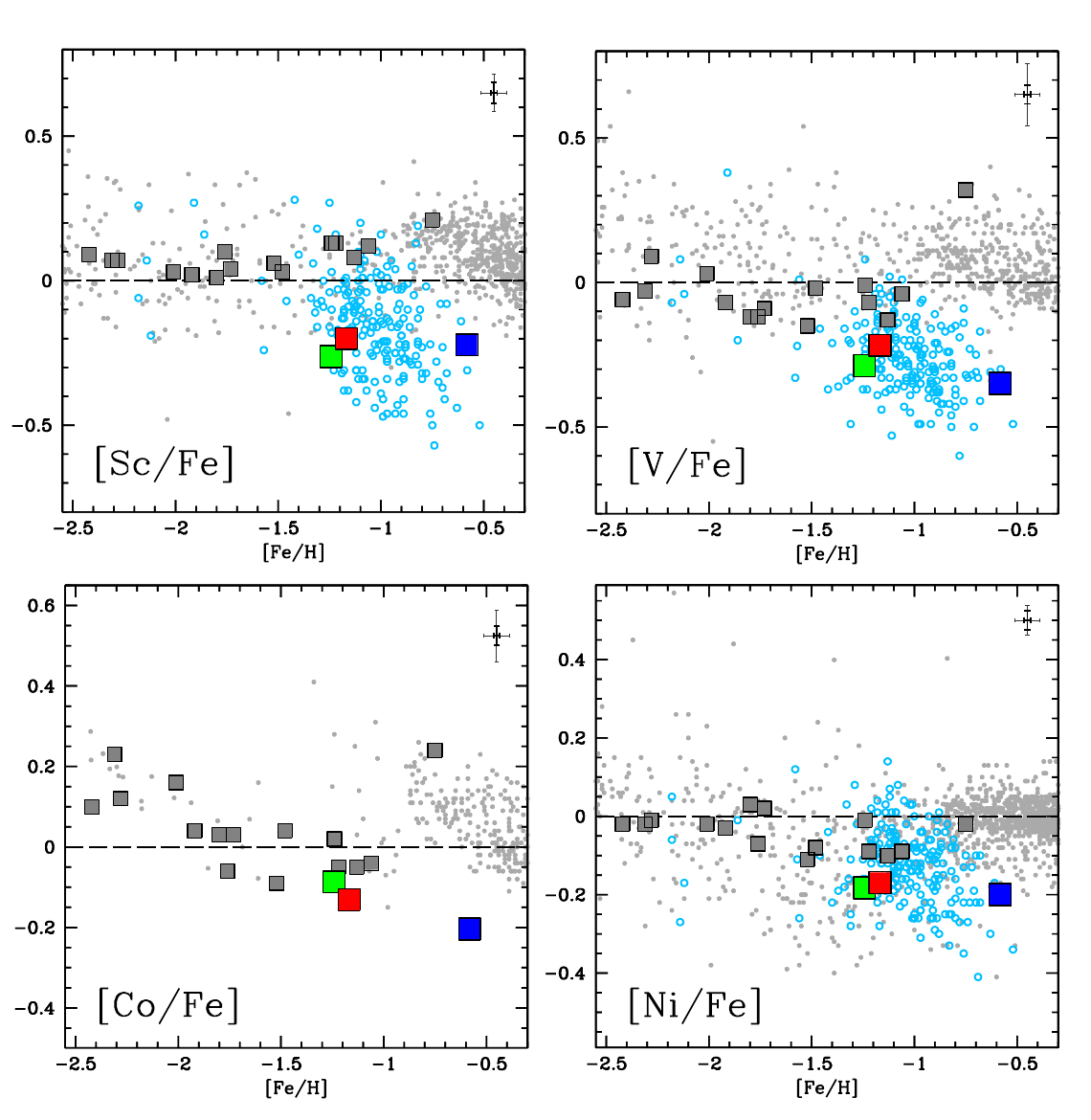}
\includegraphics[scale=0.45]{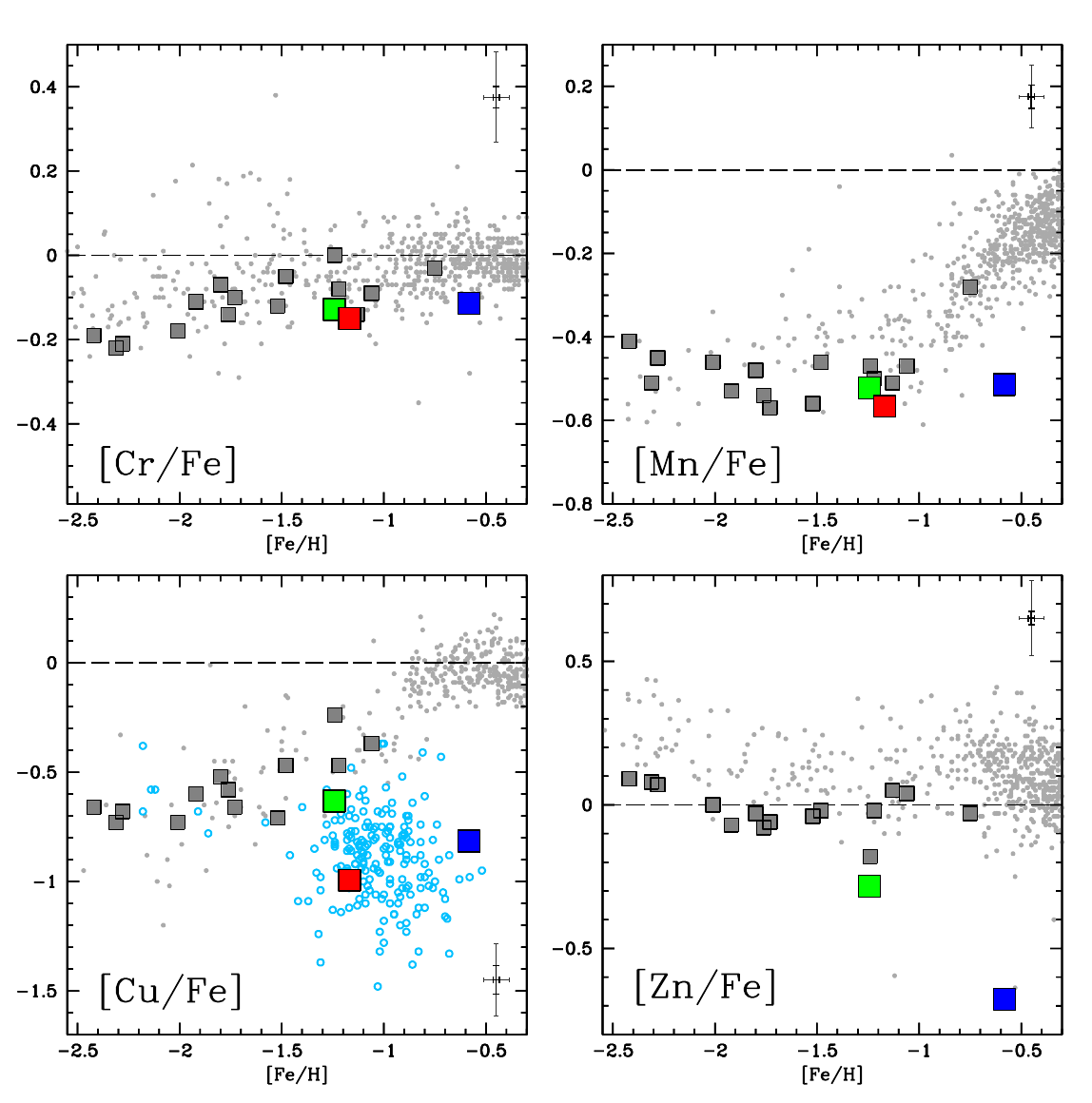}
\caption{Iron-peak elements abundance ratios as a function of [Fe/H], for SMC and MW GCs and field stars. 
Same symbols and data as in figure \ref{alfa_el_smc}. 
MW field stars data are from \cite{Fulbright2000, Stephens2002, Gratton2003, Reddy2003, 
Reddy2006, Bensby2005, Adibekyan2012, Roederer2012, Reggiani2017}.}
\label{iron1_el_smc}
\end{figure*}

\subsection{Slow neutron-capture elements}

\begin{figure*}
\centering
\includegraphics[scale=0.75]{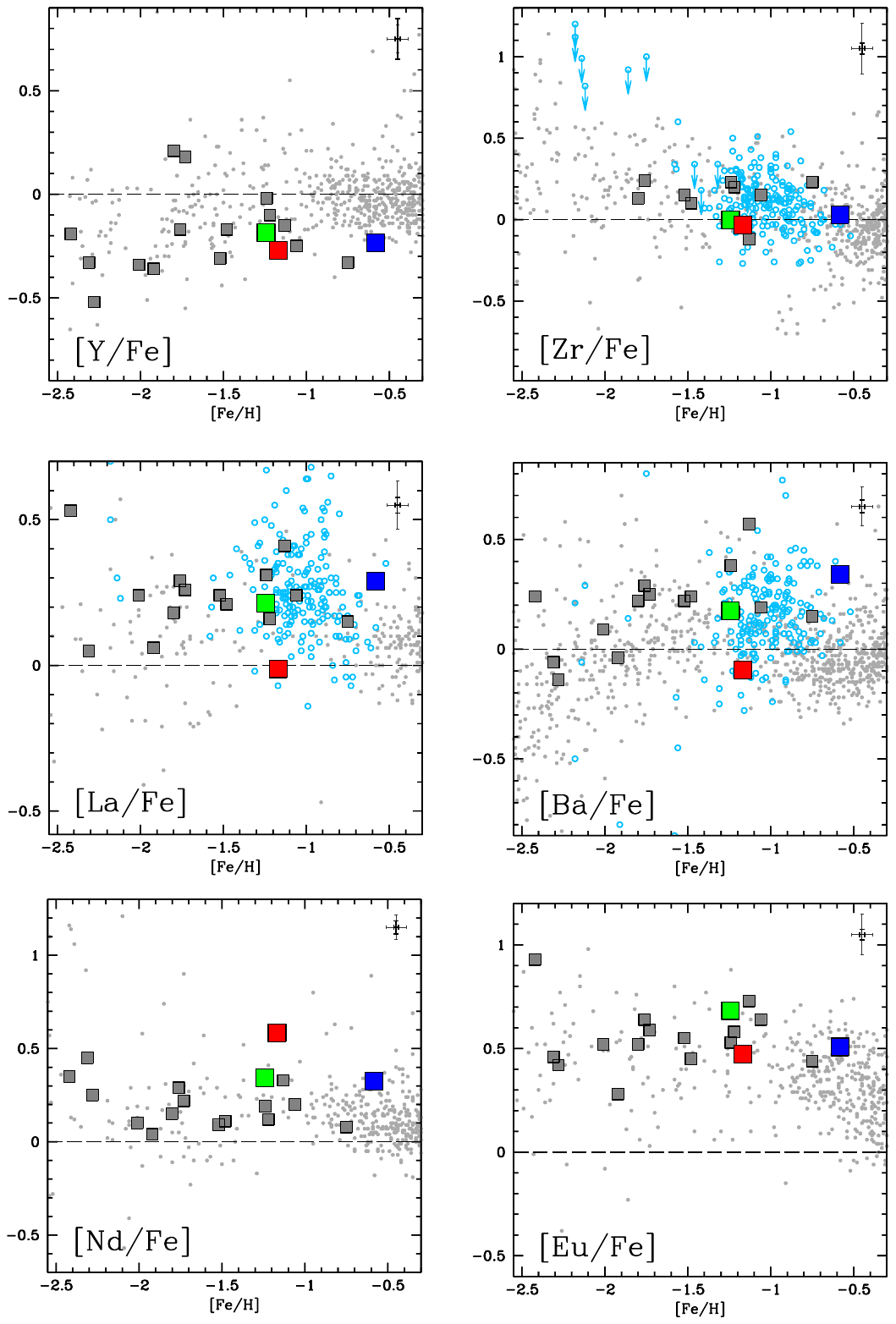}
\caption{Neutron-capture elements abundance ratios (Y, Zr, Ba, La, Nd and Eu, from 
top left panel to bottom right panel, respectively) as a function of [Fe/H], for SMC and 
MW GCs and field stars. Same symbols and data as in figure \ref{alfa_el_smc}. 
MW field stars data are from \cite{Edvardsson1993, Burris2000, Fulbright2000, Stephens2002, 
Reddy2003, Reddy2006, Barklem2005, Bensby2005, Roederer2012, mishenina2013, Yan2015, Battistini2016, 
Reggiani2017, Forsberg2019}.}
\label{neutron_el_smc}
\end{figure*}

The derived abundance ratios for Y, Zr, Ba, La and Nd are reported in Fig.~\ref{neutron_el_smc} 
as a function of the metallicity. 
These elements are mainly produced through s-processes in Asymptotic Giant Branch (AGB) stars.
In particular the light s-process elements Y and Zr are produced by AGB stars over a large mass range, 
while the heavy s-process elements Ba, La and Nd 
are mainly produced by AGB stars with masses lower than 4 $M_{\odot}$. 
The ratio of heavy to light s-process elements is sensitive to the metallicity 
\citep{Gallino98,Busso1999}: increasing the metallicity of the AGB stars, 
the production of light s-process elements is favoured against the heavy ones.

The SMC GCs have similar [Y/Fe] and [Zr/Fe], consistent with the MW stars, suggesting 
a similar enrichment by high-metallicity and/or intermediate-mass AGB stars. 
Interesting differences are observed for Ba and La 
(Nd is known to be equally produced by s- and r-processes).
In fact, NGC~121 has [Ba/Fe] and [La/Fe] slightly lower than the MW GCs, while 
NGC~339 has [Ba/Fe] and [La/Fe] ratios higher by $\sim$0.3 dex than MW clusters at similar [Fe/H]. 
NGC~419 has enhanced values for both the abundance ratios.

We use the [(Ba+La)/(Y+Zr)] abundance ratio to evaluate the relative contribution 
of heavy to light s-process elements. This ratio increases with decreasing the cluster age, 
from $\sim$+0.1 dex in NGC~121 up to +0.4 dex in NGC~419. This indicates that 
at younger ages the production of these elements is dominated by metal-poor AGB stars 
that produce especially heavy s-process elements, as expected.

\subsection{Rapid neutron-capture elements}
The r-processes occur in the presence of a very high neutron flux ($>10^{20}$ neutron/${\rm cm}^{3}$). 
Several sites of production have been proposed, as neutron star mergers \citep{lattimer74}, 
collapsars \citep{Siegel2019} and proto-magnetars \citep{nishimura15}. 
The detection of r-process elements in the kilonovae event associated to the gravitational wave 
GW170817 \citep{abbott17} confirmed that neutron star mergers are an important r-process production site, 
even if the time delay distribution for the merging process remains unknown.
However, chemical evolution models including neutron star mergers as 
the only r-process source are not able to reproduce the Galactic [Eu/Fe] distribution \citep[see e.g.][]{molero21}.

We derived abundances for the pure r-process element Eu. 
The only measures of Eu abundances in the SMC available  so far are provided by \citet{reggiani21}, 
who measured [Eu/Fe] in two metal-poor ([Fe/H]$<$--2.0 dex) SMC field stars. 
These two stars are characterised by high value of [Eu/Fe] ($\sim$+0.8/+1.0 dex), similar to those 
measured in LMC and MW stars of similar metallicity. 
The three SMC GCs discussed here display enhanced [Eu/Fe] values compatible with those of the MW 
GCs control sample. This finding suggests that the production of r-process is very efficient 
in the SMC over a large range of ages, in particular it continues also at an age of $\sim$1.5 Gyr
(the age of NGC~419) where the contribution of SNe Ia is significant, as demonstrated by its low 
[$\alpha$/Fe] ratios.
The consistent [Eu/Fe] at various [Fe/H] ratios seems to indicate that the timescales for the synthesis 
of Eu and Fe (the production of the latter being dominated by SNe~Ia), are fairly similar. 
The measured [Eu/Fe] ratios thus provide a crucial constraint to the timescale of Eu production 
and seem to suggest a dominant contribution from neutron star mergers. 
This issue will be further analysed by us by means of detailed chemical evolution models (D. Romano et al., in prep.).

\subsection{Relative contribution of r- and s-processes}
The simultaneous measure of s-process elements produced by AGB stars and the pure r-process element Eu 
allows us to study the contribution of the r-process at different ages.
In fact, because elements like Ba and La are mainly produced by AGB stars, this contribution 
starts to dominate the enrichment with a delay after the onset of the SF. Therefore, at low metallicity, 
the production of Ba and La is dominated by the r-processes. 
The ratio [(Ba+La)/Eu] provides the relative contribution of these two processes. 
NGC~121 and NGC~339 have similar [(Ba+La)/Eu] ratios ($\sim$--0.5 dex), close to the 
theoretical values predicted in the case of pure r-process \citep[see e.g.][]{bisterzo14}. 
This finding suggests that the production of Ba and La in these clusters 
is still dominated by the r-processes and the contribution by AGB stars is negligible.
However, in NGC~419, this ratio climbs to -0.2 dex, indicating a rising contribution from AGB stars. 
A similar increase of the ratio between s-process and Eu 
with the metallicity has been observed in other dwarf galaxies \citep[see e.g.][]{vds13,mcw13,hill19,Minelli2021}.

\section{Comparison between NGC~121, Milky Way in-situ and Gaia-Enceladus clusters}
\label{comparison}

The old cluster NGC~121 allows us to compare the chemical properties of the SMC in its early stages 
with those of MW GCs of similar ages.
The SMC reached an iron content of [Fe/H]$\sim$--1.2 dex at an age of $\sim$10.5 Gyr. 
We compare the properties (age, [Fe/H] and [$\alpha$/Fe]) of NGC~121 with those of the MW in-situ GCs and 
those associated to the main marger event of the MW, named Gaia-Enceladus \citep[G-E,][]{helmi18}. 
It has been suggested that the stellar mass of the progenitor of G-E should be comparable with the present-day stellar 
mass of the SMC, $\sim5-6\times 10^{8} M_{\odot}$ \citep{vdm09,helmi18}.

\begin{figure}
\includegraphics[width=\hsize,clip=true]{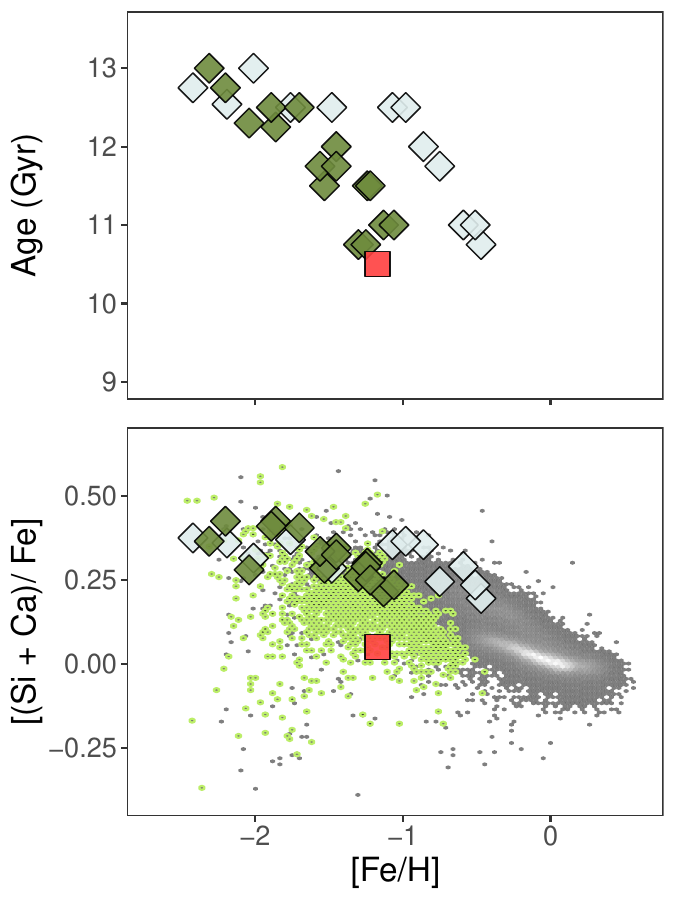}
\caption{Behaviour of age and [(Si+Ca)/Fe] as a function of [Fe/H] 
(upper and lower panel, respectively) for the MW in-situ GCs 
(grey diamonds), from Gaia-Enceladus (green diamonds) and for NGC~121 
(red square). For MW GCs ages are from \citet{vdb13}. The chemical abundances are 
from the MW control sample of this study and from 
\citet{marino11,carretta13,munoz13,carretta14,koch14,marino15,sanroman15,
johnson16,rojas16,villanova16,massari17,murag18,masseron19,koch21,marino21}. 
Dynamical association of the GCs are from \citet{massari19}.
In the lower panel, [(Si+Ca)/Fe] from the APOGEE results from 17th Data Release of the Sloan Digital Sky Survey 
\citep{sdss22} are shown for the sake of comparison, considering MW giant stars within 2 kpc (grey points) and Gaia-Enceladus 
stars (light green circles) selected according to \citet{hasselquist21}.}
\label{aaf}
\end{figure}

We consider the trends of age versus metallicity for the MW in-situ and G-E clusters 
adopting the dynamical selection by \citet{massari19}. 
The upper panel of Fig.~\ref{aaf} shows the behaviour of age from \citet{vdb13} as a function of [Fe/H] for the MW 
in-situ and G-E clusters, in comparison with NGC~121. 
For [Fe/H] $>$--1.5 dex, the MW in-situ and G-E clusters define two well-distinct sequences in the age-[Fe/H] plane. 
It is remarkable that the MW in-situ GCs at a comparable age of that of NGC~121 are 
significantly more metal-rich, reaching [Fe/H]$\sim$--0.5 dex. 
On the other hand, NGC~121 well matches with the AMR described by G-E GCs. 
Yet, despite the consistency of NGC~121 with the G-E AMR might suggest that its former progenitor (the SMC) 
followed an evolution similar to that of G-E, a significant difference arises 
when considering their $\alpha$ elements, in the form of [(Si+Ca)/Fe] ratios 
(we consider only these two $\alpha$-elements because they are available for a large number of GCs, 
while Ti abundances are not available for some of them). 
The lower panel of Fig.~\ref{aaf} shows the behaviour of [(Si+Ca)/Fe]  as a function of [Fe/H] 
for MW in-situ, G-E clusters and for NGC~121.
The clusters associated to G-E, despite starting to deviate from the sequence of the MW in-situ GCs at
lower [Fe/H], still show values of [(Si+Ca)/Fe]$\sim$0.2 dex at the metallicity of NGC~121. 
This cluster, instead, shows a nearly solar value, indicating
a formation environment already dominated by SNIa enrichment, at variance with
the G-E clusters for which SN~Ia have just started to pollute their CC-SNe dominated star-forming gas.
This comparison provides robust evidence for distinct chemical evolution histories between the SMC and GE, with the SMC having 
experienced a slower and less efficient star formation compared to G-E. 
This is consistent with the fact that the SMC mass at $\sim$10.5 Gyr was likely significantly lower than that of G-E.
A similar result has been suggested also by \citet{hasselquist21} and \citet{fernandes23} 
based on the chemical patterns of SMC and G-E field stars (hence without taking the stellar 
ages into account, which can be done via GCs). 
It should be emphasised that the interaction of G-E with the MW during the cannibalisation process 
might have triggered a SF burst and, hence, higher SF rates than experienced by the SMC during its quiet 
evolution in isolation. A sudden SF burst is expected to restore the [$\alpha$/Fe] ratios 
to values more typical of those of CC-SN ejecta.

\section{Summary}
\label{summary}

This study provides a time-resolved reconstruction of the chemical enrichment history of the SMC 
galaxy by presenting for the first time the detailed chemical composition based on high-resolution 
spectra of three clusters in a wide range interval.
The following is a summary of the main findings:
\begin{itemize}
\item 
We can place constraints on the metallicity that the SMC attained at certain ages 
using the derived [Fe/H] of the three GCs. The relationship between [Fe/H] and cluster age is 
in excellent accord with the theoretical AMR proposed by \citet{Pagel1998}, which incorporated a sizeable 
halt in star formation between the first burst and the subsequent one at 4 Gyr.
However, given the wide metallicity variance seen at any age \citep[see e.g.][]{Parisi2022}, our sample of clusters should
not be seen as being as indicative of the entire SMC.
This spread may have been influenced by the SMC's metallicity gradient and the existence of 
varied chemical enrichment histories in different SMC sites (as originally mentioned in Paper I), 
which reflect turbulent interactions between the two Clouds.
\item
We derived the abundance ratios for elements belonging to the main groups ($\alpha$, 
iron-peak and neutron-capture) and we compared them with 
the abundances of SMC field stars (Paper I). 
The two samples have similar (homogenous) abundances in all the elements, indicating that GCs and field stars 
experienced a similar chemical enrichment.
The comparison between abundances of SMC and MW GCs highlights their different chemical enrichment histories. 
In particular, the enrichment of the SMC (at least in the age range explored in this study) 
is dominated by SN~Ia and displays a lower contribution by massive stars.
\item 
Despite having an age difference of about 4-5 Gyr, the ancient cluster NGC~121 and the intermediate-age cluster NGC~339 
exhibit identical [Fe/H] values and similar abundances for the majority of the measured elements. 
Differences of about 0.2-0.3 dex are observed only 
for heavy s-process elements. These results suggest a similar chemical enrichment by CC-SNe and SN~Ia but an increase 
of the contribution of the metal-poor AGB stars with the age.
\item 
The young (1.4 Gyr) cluster NGC~419 has chemical patterns that are markedly different from those of the other GCs, 
including higher [Fe/H], lower [$\alpha$/Fe] and [Zn/Fe] ratios, and higher heavy s-process elements. 
These abundances show that the metal-poor AGB stars and SN Ia made a significant contribution to the evolution 
of the SMC at that age, although HNe made a less significant contribution, as shown by the extremely low [Zn/Fe] ratio. 
It is interesting to note that [Eu/Fe] in this cluster is enhanced, indicating that the production of r-process in the 
SMC is extremely efficient throughout a wide range of ages, including at an age of around 1.5 Gyr where 
the contribution of SNe~Ia is considerable as evidenced by its low [$\alpha$/Fe] ratios.

\item 
We contrasted the characteristics of the old SMC cluster NGC 121 with those of MW clusters formed 
in-situ and clusters connected  to the G-E merger event. The early SMC and G-E have both attained 
the same metallicity at the same age as NGC 121, but their [$\alpha$/Fe] ratios are different, with the latter 
displaying enhanced values in contrast to NGC 121, which has solar-scaled [$\alpha$/Fe]  ratios. This shows that the SMC 
and G-E have distinct histories of chemical enrichment, 
most likely a quiet and a bursty one, respectively, because of the interaction of G-E with the MW as opposed 
to the evolution in isolation of the SMC. Furthermore, the SMC had at early ages had a mass lower than G-E.

\end{itemize}

\begin{acknowledgements}
We thanks the anonymous referee for the useful and constructive suggestions.
This research is funded  by the project "Light-on-Dark" , granted by the Italian MIUR
through contract PRIN-2017K7REXT.
C. Lardo acknowledges funding from Ministero dell'Università e della Ricerca through the 
Programme {\em Rita Levi Montalcini} (grant PGR18YRML1).

\end{acknowledgements}

\begin{appendix}
\section{Chemical abundance ratios for the MW control sample}
\label{app}

The MW control sample includes spectra of giant stars of 16 GCs retrieved from the ESO archive. 
All the spectra have been secured with UVES-FLAMES adopting the Red Arm 580 configuration.
We refer to \citet{Mucciarelli2021} for further details on the MW GC sample 
(i.e. the number of observed stars and the corresponding observative programs) and its analysis. 

In tables~\ref{mw1}, ~\ref{mw2}, ~\ref{mw3} and \ref{mw4} we list for each MW target cluster the average abundance ratios together 
with the corresponding standard error and the observed dispersion of the mean, $\sigma_{obs}$.

\begin{table*}[!h]\tiny
\caption{Milky Way globular clusters: average abundance ratios of iron and $\alpha$ elements (Si, Ca, Ti~I and Ti~II) 
together with the corresponding standard error and dispersion of the mean.}
\label{mw1}
\begin{tabular}{c cc cc cc cc cc}
{\rm Cluster} &  ${\rm [Fe/H]}$  & $\sigma_{obs}$  
&  ${\rm [Si/Fe]}$  & $\sigma_{obs}$ 
&  ${\rm [Ca/Fe]}$  & $\sigma_{obs}$ 
&  ${\rm [Ti/Fe]}$  & $\sigma_{obs}$ 
&  ${\rm [Ti~II/Fe]}$  & $\sigma_{obs}$\\
\hline 
NGC~104   &  --0.75$\pm$0.01    &   0.03        &   +0.28$\pm$0.01    & 0.03   &	+0.21$\pm$0.02  &  0.07   &  +0.33$\pm$0.01    &  0.03  & +0.18$\pm$0.01  &  0.04    \\
NGC~288   &  --1.24$\pm$0.01    &   0.04	    &   +0.33$\pm$0.01    & 0.03   &	+0.27$\pm$0.01  &  0.03   &  +0.24$\pm$0.01    &  0.03  & +0.26$\pm$0.01  &  0.02	\\ 
NGC~1851  &  --1.13$\pm$0.01    &   0.04	    &   +0.25$\pm$0.01    & 0.05   &	+0.18$\pm$0.01  &  0.05   &  +0.09$\pm$0.01    &  0.04  & +0.25$\pm$0.01  &  0.04    \\ 
NGC~2808  &  --1.06$\pm$0.02    &   0.07        &   +0.26$\pm$0.01    & 0.04   &	+0.21$\pm$0.01  &  0.02   &  +0.14$\pm$0.01    &  0.03  & +0.18$\pm$0.01  &  0.02    \\ 
NGC~4590  &  --2.28$\pm$0.01    &   0.05	    &   +0.35$\pm$0.04    & 0.06   &	+0.23$\pm$0.01  &  0.04   &  +0.16$\pm$0.01    &  0.04  & +0.23$\pm$0.01  &  0.04	 \\ 
NGC~5634  &  --1.80$\pm$0.02    &   0.05	    &   +0.29$\pm$0.01    & 0.04   &	+0.22$\pm$0.01  &  0.03   &  +0.21$\pm$0.01    &  0.04  & +0.35$\pm$0.03  &  0.07	\\ 
NGC~5824  &  --1.92$\pm$0.02    &   0.04	    &   +0.36$\pm$0.03    & 0.08   &	+0.24$\pm$0.01  &  0.02   &  +0.18$\pm$0.02    &  0.05  & +0.19$\pm$0.02  &  0.05	\\ 
NGC~5904  &  --1.22$\pm$0.01    &   0.03	    &   +0.29$\pm$0.01    & 0.03   &	+0.21$\pm$0.01  &  0.03   &  +0.15$\pm$0.01    &  0.03  & +0.25$\pm$0.01  &  0.05	 \\ 
NGC~6093  &  --1.76$\pm$0.01    &   0.03	    &   +0.35$\pm$0.01    & 0.04   &	+0.28$\pm$0.01  &  0.03   &  +0.15$\pm$0.01    &  0.04  & +0.34$\pm$0.01  &  0.04	 \\ 
NGC~6397  &  --2.01$\pm$0.01    &   0.03	    &   +0.37$\pm$0.02    & 0.08   &	+0.26$\pm$0.01  &  0.03   &  +0.19$\pm$0.01    &  0.02  & +0.22$\pm$0.01  &  0.04	 \\ 
NGC~6752  &  --1.48$\pm$0.01    &   0.03	    &   +0.29$\pm$0.01    & 0.03   &	+0.28$\pm$0.01  &  0.02   &  +0.24$\pm$0.01    &  0.03  & +0.22$\pm$0.01  &  0.05	 \\ 
NGC~6809  &  --1.73$\pm$0.01    &   0.03        &   +0.26$\pm$0.01    & 0.04   &	+0.25$\pm$0.01  &  0.03   &  +0.15$\pm$0.01    &  0.02  & +0.28$\pm$0.01  &  0.03	\\ 
NGC~7078  &  --2.42$\pm$0.02    &   0.07	    &   +0.47$\pm$0.03    & 0.08   &	+0.28$\pm$0.01  &  0.02   &  +0.23$\pm$0.02    &  0.06  & +0.29$\pm$0.02  &  0.05	 \\
NGC~7099  &  --2.31$\pm$0.01    &   0.05	    &   +0.45$\pm$0.01    & 0.02   &	+0.28$\pm$0.01  &  0.03   &  +0.20$\pm$0.01    &  0.05  & +0.26$\pm$0.01  &  0.04	 \\

\hline							  
\hline							  
\end{tabular}
\end{table*}


\begin{table*}\tiny
\caption{Milky Way globular clusters: average abundance ratios of iron-peak elements (Sc, V, Cr and Mn)
together with the corresponding standard error and dispersion of the mean.}
\label{mw2}
\begin{tabular}{c cc cc cc cc }
{\rm Cluster} 
&  ${\rm [Sc~II/Fe]}$  & $\sigma_{obs}$ 
&  ${\rm [V/Fe]}$  & $\sigma_{obs}$ 
&  ${\rm [Cr/Fe]}$  & $\sigma_{obs}$ 
&  ${\rm [Mn/Fe]}$  & $\sigma_{obs}$  \\
\hline 

NGC~104   &   +0.21$\pm$0.02	& 0.06    &    +0.32$\pm$0.02  &  0.07   &   --0.03$\pm$0.02  &  0,07	&  --0.28$\pm$0.01   &	0.04   \\
NGC~288   &   +0.13$\pm$0.01	& 0.03    &   --0.01$\pm$0.02  &  0.06   &    +0.00$\pm$0.01  &  0.03	&  --0.47$\pm$0.01   &	0.03   \\ 
NGC~1851  &   +0.08$\pm$0.01	& 0.06    &   --0.13$\pm$0.02  &  0.08   &   --0.14$\pm$0.01  &  0.06	&  --0.51$\pm$0.02   &	0.04  \\ 
NGC~1904  &   +0.06$\pm$0.01	& 0.03    &   --0.15$\pm$0.01  &  0.05   &   --0.12$\pm$0.01  &  0.03	&  --0.56$\pm$0.01   &	0.05   \\ 
NGC~2808  &   +0.12$\pm$0.01	& 0.06    &   --0.04$\pm$0.04  &  0.10   &   --0.09$\pm$0.01  &  0.05	&  --0.47$\pm$0.01   &	0.03   \\ 
NGC~4590  &   +0.07$\pm$0.01	& 0.04    &    +0.09$\pm$0.02  &  0.03   &   --0.21$\pm$0.01  &  0.07	&  --0.45$\pm$0.01   &	0.04   \\ 
NGC~5634  &   +0.01$\pm$0.03	& 0.07    &   --0.12$\pm$0.03  &  0.06   &   --0.07$\pm$0.01  &  0.04	&  --0.48$\pm$0.03   &	0.07   \\ 
NGC~5824  &   +0.02$\pm$0.03	& 0.06    &   --0.07$\pm$0.03  &  0.06   &   --0.11$\pm$0.01  &  0.01	&  --0.53$\pm$0.03   &	0.07   \\ 
NGC~5904  &   +0.13$\pm$0.01	& 0.03    &   --0.07$\pm$0.01  &  0.04   &   --0.08$\pm$0.01  &  0.05	&  --0.50$\pm$0.01   &	0.03  \\ 
NGC~6093  &   +0.10$\pm$0.01	& 0.04    &   --0.12$\pm$0.01  &  0.04   &   --0.14$\pm$0.01  &  0.04	&  --0.54$\pm$0.01   &	0.02   \\ 
NGC~6397  &   +0.03$\pm$0.01	& 0.04    &    +0.03$\pm$0.02  &  0.07   &   --0.18$\pm$0.01  &  0.03	&  --0.46$\pm$0.03   &	0.10   \\ 
NGC~6752  &   +0.03$\pm$0.01	& 0.03    &   --0.02$\pm$0.01  &  0.04   &   --0.05$\pm$0.01  &  0.03	&  --0.46$\pm$0.01   &	0.03   \\ 
NGC~6809  &   +0.04$\pm$0.01	& 0.04    &   --0.09$\pm$0.01  &  0.03   &   --0.10$\pm$0.01  &  0.04	&  --0.57$\pm$0.01   &	0.05  \\ 
NGC~7078  &   +0.09$\pm$0.02	& 0.07    &   --0.06$\pm$0.06  &  0.10   &   --0.19$\pm$0.01  &  0.05	&  --0.41$\pm$0.03   &	0.07   \\
NGC~7099  &   +0.07$\pm$0.01	& 0.04    &   --0.03$\pm$0.02  &  0.05   &   --0.22$\pm$0.01  &  0.03	&  --0.51$\pm$0.02   &	0.06   \\

\hline					     		  
\hline							  
\end{tabular}
\end{table*}


\begin{table*}\tiny
\caption{Milky Way globular clusters: average abundance ratios of iron-peak elements (Co, Ni, Cu and Zn)
together with the corresponding standard error and dispersion of the mean.}
\label{mw3}
\begin{tabular}{c cc cc cc cc }
{\rm Cluster} 
&  ${\rm [Co/Fe]}$  & $\sigma_{obs}$
&  ${\rm [Ni/Fe]}$  & $\sigma_{obs}$ 
&  ${\rm [Cu/Fe]}$  & $\sigma_{obs}$
&  ${\rm [Zn/Fe]}$  & $\sigma_{obs}$ \\
\hline 

NGC~104   &     +0.24$\pm$0.01    &   0.04   &   --0.02$\pm$0.01  &  0.02   &   ---              &   ---       &  --0.03$\pm$0.03	 &    0.09  \\
NGC~288   &     +0.02$\pm$0.01    &   0.03   &   --0.01$\pm$0.01  &  0.03   & --0.24$\pm$0.02    &   0.05      &  --0.18$\pm$0.04	 &    0.14   \\ 
NGC~1851  &    --0.05$\pm$0.01    &   0.04   &   --0.10$\pm$0.01  &  0.02   &   ---              &   ---       &   +0.05$\pm$0.03	 &    0.14  \\ 
NGC~1904  &    --0.09$\pm$0.02    &   0.07   &   --0.11$\pm$0.01  &  0.02   & --0.71$\pm$0.01    &   0.04      &  --0.04$\pm$0.02	 &    0.14\\ 
NGC~2808  &    --0.04$\pm$0.01    &   0.04   &   --0.09$\pm$0.01  &  0.03   & --0.37$\pm$0.02    &   0.12      &   +0.04$\pm$0.05	 &    0.17 \\ 
NGC~4590  &     +0.12$\pm$0.02    &   0.03   &   --0.01$\pm$0.01  &  0.05   & --0.68$\pm$0.02    &   o.04      &   +0.07$\pm$0.03	 &    0.10 \\ 
NGC~5634  &     +0.03$\pm$0.01    &   0.04   &    +0.03$\pm$0.01  &  0.04   & --0.52$\pm$0.04    &   0.11      &  --0.03$\pm$0.05	 &    0.15 \\ 
NGC~5824  &     +0.04$\pm$0.03    &   0.08   &   --0.03$\pm$0.01  &  0.02   & --0.60$\pm$0.04    &   0.11      &  --0.07$\pm$0.03	 &    0.07 \\ 
NGC~5904  &    --0.05$\pm$0.01    &   0.03   &   --0.09$\pm$0.01  &  0.02   & --0.47$\pm$0.02    &   0.06      &  --0.02$\pm$0.02	 &    0.09 \\ 
NGC~6093  &    --0.06$\pm$0.01    &   0.03   &   --0.07$\pm$0.01  &  0.02   & --0.58$\pm$0.01    &   0.03      &  --0.08$\pm$0.02	 &    0.07 \\ 
NGC~6397  &     +0.16$\pm$0.02    &   0.04   &   --0.02$\pm$0.01  &  0.03   & --0.73$\pm$0.04    &   0.09      &   +0.00$\pm$0.02	 &    0.06 \\ 
NGC~6752  &     +0.04$\pm$0.01    &   0.03   &   --0.08$\pm$0.01  &  0.02   & --0.47$\pm$0.01    &   0.06      &  --0.02$\pm$0.03	 &    0.12\\ 
NGC~6809  &     +0.03$\pm$0.02    &   0.06   &    +0.00$\pm$0.01  &  0.02   & --0.66$\pm$0.01    &   0.05      &  --0.06$\pm$0.01	 &    0.05 \\ 
NGC~7078  &     +0.10$\pm$0.01    &   0.02   &   --0.02$\pm$0.02  &  0.06   & --0.66$\pm$0.03    &   0.07      &   +0.09$\pm$0.03	 &    0.12 \\
NGC~7099  &     +0.23$\pm$0.03    &   0.08   &   --0.02$\pm$0.01  &  0.03   & --0.73$\pm$0.03    &   0.10      &   +0.08$\pm$0.02	 &    0.08 \\

\hline					     		  
\hline							  
\end{tabular}
\end{table*}


\begin{table*}\tiny
\caption{Milky Way globular clusters: average abundance ratios of neutron-capture elements (Y, Zr, Ba, La, Nd and Eu) 
together with the corresponding standard error and dispersion of the mean.}
\label{mw4}
\begin{tabular}{c cc cc cc cc cc cc }
{\rm Cluster}  
&  ${\rm [Y~II/Fe]}$  & $\sigma_{obs}$ 
&  ${\rm [Zr/Fe]}$  & $\sigma_{obs}$ 
&  ${\rm [Ba~II/Fe]}$  & $\sigma_{obs}$ 
&  ${\rm [La~II/Fe]}$  & $\sigma_{obs}$
&  ${\rm [Nd~II/Fe]}$  & $\sigma_{obs}$
&  ${\rm [Eu~II/Fe]}$  & $\sigma_{obs}$\\
\hline 

NGC~104        &  --0.33$\pm$0.04  &	0.11   & +0.23$\pm$0.05  & 0.15      &  +0.15$\pm$0.01    &  0.03 &   +0.15$\pm$0.02    &  0.07 & +0.08$\pm$0.01    &  0.05 &   +0.44$\pm$0.01    & 0.04     \\
NGC~288        &  --0.02$\pm$0.03  &	0.10   & +0.23$\pm$0.03  & 0.14	     &  +0.38$\pm$0.02    &  0.07 &   +0.31$\pm$0.01    &  0.02 & +0.19$\pm$0.01    &  0.03 &   +0.53$\pm$0.02    & 0.06     \\ 
NGC~1851       &  --0.15$\pm$0.03  &	0.13   &--0.12$\pm$0.03  & 0.13	     &  +0.57$\pm$0.03    &  0.12 &   +0.41$\pm$0.03    &  0.17	& +0.33$\pm$0.02    &  0.12 &   +0.73$\pm$0.01    & 0.05  \\ 
NGC~1904       &  --0.31$\pm$0.02  &	0.05   & +0.15$\pm$0.04  & 0.13      &  +0.22$\pm$0.02    &  0.05 &   +0.24$\pm$0.03    &  0.09 & +0.09$\pm$0.02    &  0.07 &   +0.55$\pm$0.04    & 0.10       \\ 
NGC~2808       &  --0.25$\pm$0.03  &	0.11   & +0.15$\pm$0.06  & 0.15	     &  +0.19$\pm$0.03    &  0.10 &   +0.24$\pm$0.01	&  0.07 & +0.20$\pm$0.02    &  0.07 &   +0.64$\pm$0.02    & 0.08      \\ 
NGC~4590       &  --0.52$\pm$0.02  &	0.06   &   ---           &   ---     & --0.14$\pm$0.02    &  0.09 &     ---             &  ---  & +0.25$\pm$0.02    &  0.05 &   +0.42$\pm$0.01    & 0.04   \\ 
NGC~5634       &   +0.21$\pm$0.06  &    0.17   & +0.13$\pm$0.09  & 0.21	     &  +0.22$\pm$0.07    &  0.19 &   +0.18$\pm$0.03    &  0.07	& +0.15$\pm$0.03    &  0.08 &   +0.52$\pm$0.03    & 0.07     \\ 
NGC~5824       &  --0.36$\pm$0.01  &	0.03   &  ---            &   ---     & --0.04$\pm$0.05    &  0.10 &   +0.06$\pm$0.03    &  0.07	& +0.04$\pm$0.04    &  0.12 &   +0.28$\pm$0.03    & 0.07   \\ 
NGC~5904       &  --0.10$\pm$0.06  &	0.25   & +0.20$\pm$0.05  & 0.21	     &  +0.17$\pm$0.02    &  0.07 &   +0.16$\pm$0.02    &  0.07	& +0.12$\pm$0.01    &  0.04 &   +0.58$\pm$0.02    & 0.07    \\ 
NGC~6093       &  --0.17$\pm$0.03  &	0.11   & +0.24$\pm$0.04  & 0.11	     &  +0.29$\pm$0.05    &  0.14 &   +0.29$\pm$0.05	&  0.12 & +0.29$\pm$0.04    &  0.12 &	+0.64$\pm$0.02    & 0.05     \\ 
NGC~6397       &  --0.34$\pm$0.03  &	0.10   &  ---            &   ---     &  +0.09$\pm$0.02    &  0.07 &   +0.24$\pm$0.03	&  0.08 & +0.10$\pm$0.03    &  0.10 &	+0.52$\pm$0.03    & 0.07    \\ 
NGC~6752       &  --0.17$\pm$0.02  &    0.06   & +0.10$\pm$0.04  & 0.16	     &  +0.24$\pm$0.03    &  0.10 &   +0.21$\pm$0.01	&  0.04 & +0.11$\pm$0.01    &  0.03 &	+0.45$\pm$0.01    & 0.04     \\ 
NGC~6809       &   +0.18$\pm$0.01  &	0.03   &  ---            &   ---     &  +0.25$\pm$0.03    &  0.09 &   +0.26$\pm$0.01	&  0.04 & +0.22$\pm$0.01    &  0.04 &	+0.59$\pm$0.01    & 0.05    \\ 
NGC~7078       &  --0.19$\pm$0.02  &	0.09   &  ---		     &   ---     &  +0.24$\pm$0.04    &  0.16 &   +0.53$\pm$0.06    &  0.11	& +0.35$\pm$0.02    &  0.06 &	+0.93$\pm$0.03    & 0.09    \\
NGC~7099       &  --0.33$\pm$0.02  &	0.06   &  ---		     &   ---     & --0.06$\pm$0.02    &  0.10 &   +0.05$\pm$0.02    &  0.05	& +0.45$\pm$0.02    &  0.05 &	+0.46$\pm$0.02    & 0.05     \\

\hline							  
\hline							  
\end{tabular}
\end{table*}


\newpage
\section{Comparison with APOGEE abundances}
\label{apo}

\begin{figure*}[!h]
\centering
\includegraphics[scale=0.5]{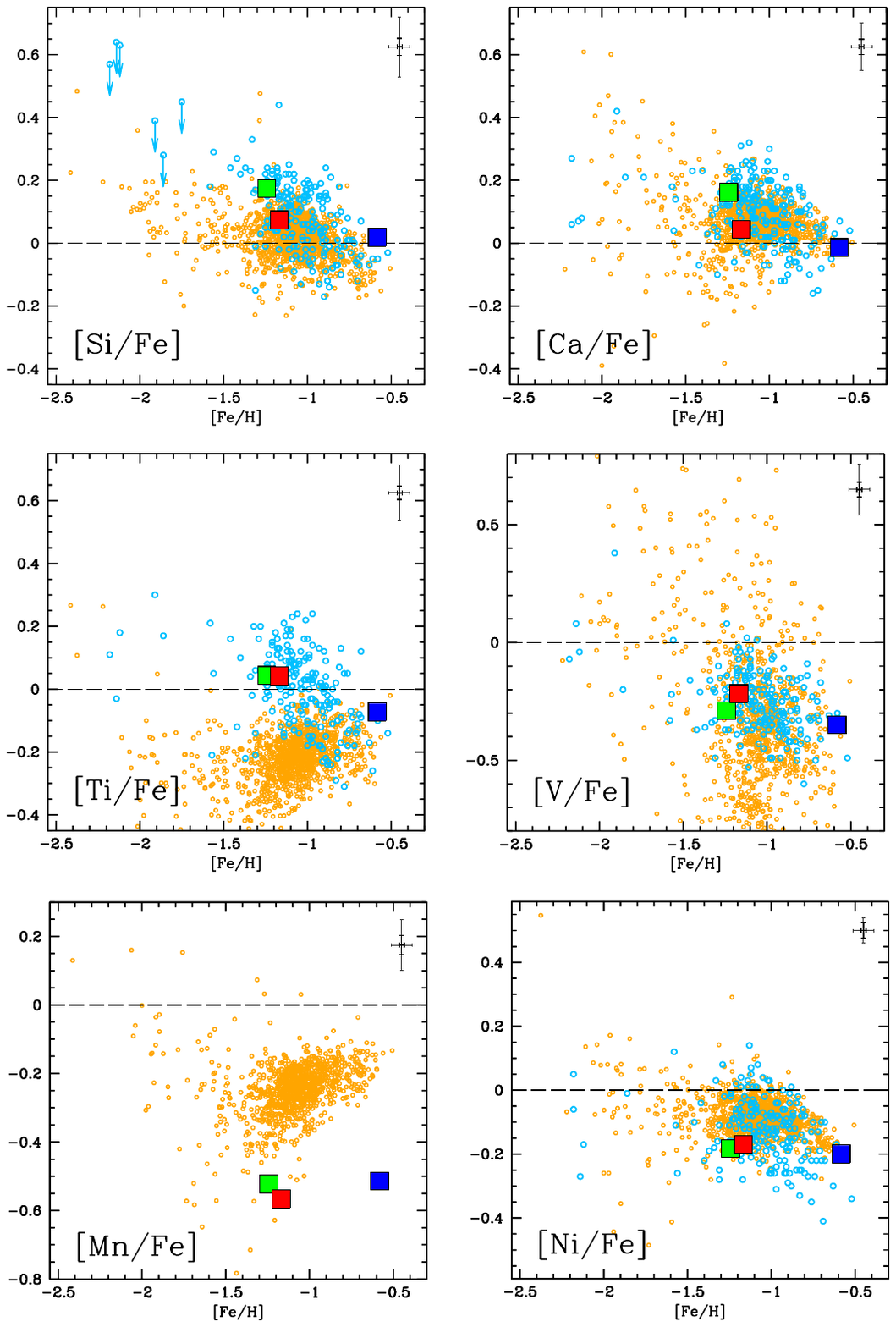}
\caption{Behaviour of [Si/Fe], [Ca/Fe], [Ti/Fe], [V/Fe], [Mn/Fe] and [Ni/Fe] 
as a function of [Fe/H] for the three SMC target clusters (squares, same colours in Fig.~\ref{cmd_gc_smc}.), 
SMC field stars of Paper I (cyan circles) and SMC field stars measured by APOGEE (orange circles).}
\label{apo_comp}
\end{figure*}

We compare the average abundance ratios for the three SMC target clusters discussed in this paper 
with the abundances derived from the APOGEE survey for SMC field stars using H-band high-resolution spectra. 
We used the results from 17th Data Release of the Sloan Digital Sky Survey 
\citep{sdss22}, adopting the selection provided by \citet{hasselquist21} for the SMC stars.
The elements in common between our analysis and APOGEE are the explosive $\alpha$-elements Si, Ca and Ti, 
and the iron-peak elements V, Mn and Ni. 
Fig.~\ref{apo_comp} shows the behaviour of the abundance ratios of these six elements as a function 
of [Fe/H] for the three SMC clusters, the SMC field stars measured in Paper I and the SMC field stars 
measured by APOGEE. 

[Si/Fe], [Ca/Fe] and [Ni/Fe] from APOGEE and FLAMES spectra agree well each other.
[Ti/Fe] measured by APOGEE is 0.2-0.3 dex lower than the values obtained with FLAMES 
and disagree with the average values of the other [$\alpha$/Fe] measured by APOGEE. 
We note as the abundances of the three explosive $\alpha$-elements measured in Paper I are in better agreement each other.
This different behaviour is reported also in the APOGEE documentation\footnote{https://www.sdss4.org/dr17/irspec/abundances/} 
where the Ti abundances in giant stars are classified as "deviant" because the measured trends
are discrepant with respect to literature expectations.
[V/Fe] exhibits a significant star-to-star scatter in the APOGEE data, larger than that of Paper I.  
According to the APOGEE documentation, V abundances in giant stars are classified as "less reliable".
Finally, [Mn/Fe] is measured by APOGEE but not in Paper I. There is an offset of $\sim$0.3 dex 
between the APOGEE [Mn/Fe] values and those measured in our SMC clusters. Mn abundances by APOGEE 
are classified as "most reliable" and we interpret this offset as due to a systematic between the optical and 
near-infrared Mn transitions.

\end{appendix}

\end{document}